\documentclass[useAMS,usenatbib]{mn2e}

\title[Galaxy Zoo: the properties of merging galaxies (I)]{Galaxy Zoo: the fraction of merging galaxies in the SDSS and their morphologies}

\author[Darg et al.]{D. W. Darg,$^1$\thanks{Email: ddarg@astro.ox.ac.uk} S. Kaviraj,$^{2,1}$\thanks{Email: skaviraj@astro.ox.ac.uk} C. J. Lintott,$^1$\thanks{Email: cjl@astro.ox.ac.uk} K. Schawinski,$^{3,4}$ M. Sarzi,$^5$ S. Bamford,$^6$ \newauthor J. Silk,$^1$ R. Proctor,$^7$ D. Andreescu,$^8$ P. Murray,$^{9}$ R. C. Nichol,$^{10}$ M. J. Raddick,$^{11}$ \newauthor A. Slosar,$^{12}$ A. S. Szalay,$^{11}$ D. Thomas,$^{10}$ J. Vandenberg.$^{11}$ \thanks{This publication has been made possible by the participation of more than 140,000 volunteers in the Galaxy Zoo project. Their contributions are individually acknowledged at \texttt{http://www.galaxyzoo.org/Volunteers.aspx}.}\\ \\
$^{1}$ Department of Physics, University of Oxford, Keble Road, Oxford, OX1 3RH, UK\\
$^{2}$ Mullard Space Science Laboratory, University College London, Holmbury St. Mary, Dorking, Surrey RH5 6NT, UK \\
$^{3}$ Department of Physics, Yale University, New Haven, CT 06511, USA\\
$^{4}$ Yale Center for Astronomy and Astrophysics, Yale University, P.O. Box 208121, New Haven, CT 06520, USA\\
$^{5}$ Centre for Astrophysics Research, University of Hertfordshire, Hatfield, AL10 9AB, UK	\\
$^{6}$ Centre for Astronomy and Particle Theory, University of Nottingham, University Park, Nottingham, NG7 2RD, UK \\
$^{7}$ Waveney Consulting/Waveney Web Services, 28 Diprose Road, Corfe Mullen, Wimborne, Dorset, BH21 3QY, UK\\
$^{8}$ LinkLab, 4506 Graystone Ave., Bronx, NY 10471, USA\\
$^{9}$ Fingerprint Digital Media, 9 Victoria Close, Newtownards, Co. Down, Northern Ireland, BT23 7GY, UK\\
$^{10}$ Institute of Cosmology and Gravitation, University of Portsmouth, Mercantile House, Hampshire Terrace, Portsmouth, PO1 2EG, UK\\
$^{11}$ Department of Physics and Astronomy, Johns Hopkins University, 3400 N. Charles St., Baltimore, MD 21218, USA\\
$^{12}$ Berkeley Center for Cosmological Physics, Lawrence Berkeley National Laboratory and Physics Department, University of California, \\Berkeley CA 94720, USA\\
}

\usepackage{graphicx}   
\usepackage{wasysym}	
\usepackage{amsfonts}	%
\usepackage{color}
\definecolor{AliceBlue}{rgb}{0.94,0.97,1.00}
\definecolor{AntiqueWhite1}{rgb}{1.00,0.94,0.86}
\definecolor{AntiqueWhite2}{rgb}{0.93,0.87,0.80}
\definecolor{AntiqueWhite3}{rgb}{0.80,0.75,0.69}
\definecolor{AntiqueWhite4}{rgb}{0.55,0.51,0.47}
\definecolor{AntiqueWhite}{rgb}{0.98,0.92,0.84}
\definecolor{BlanchedAlmond}{rgb}{1.00,0.92,0.80}
\definecolor{BlueViolet}{rgb}{0.54,0.17,0.89}
\definecolor{CadetBlue1}{rgb}{0.60,0.96,1.00}
\definecolor{CadetBlue2}{rgb}{0.56,0.90,0.93}
\definecolor{CadetBlue3}{rgb}{0.48,0.77,0.80}
\definecolor{CadetBlue4}{rgb}{0.33,0.53,0.55}
\definecolor{CadetBlue}{rgb}{0.37,0.62,0.63}
\definecolor{CornflowerBlue}{rgb}{0.39,0.58,0.93}
\definecolor{DarkBlue}{rgb}{0.00,0.00,0.55}
\definecolor{DarkCyan}{rgb}{0.00,0.55,0.55}
\definecolor{DarkGoldenrod1}{rgb}{1.00,0.73,0.06}
\definecolor{DarkGoldenrod2}{rgb}{0.93,0.68,0.05}
\definecolor{DarkGoldenrod3}{rgb}{0.80,0.58,0.05}
\definecolor{DarkGoldenrod4}{rgb}{0.55,0.40,0.03}
\definecolor{DarkGoldenrod}{rgb}{0.72,0.53,0.04}
\definecolor{DarkGray}{rgb}{0.66,0.66,0.66}
\definecolor{DarkGreen}{rgb}{0.00,0.39,0.00}
\definecolor{DarkGrey}{rgb}{0.66,0.66,0.66}
\definecolor{DarkKhaki}{rgb}{0.74,0.72,0.42}
\definecolor{DarkMagenta}{rgb}{0.55,0.00,0.55}
\definecolor{DarkOliveGreen1}{rgb}{0.79,1.00,0.44}
\definecolor{DarkOliveGreen2}{rgb}{0.74,0.93,0.41}
\definecolor{DarkOliveGreen3}{rgb}{0.64,0.80,0.35}
\definecolor{DarkOliveGreen4}{rgb}{0.43,0.55,0.24}
\definecolor{DarkOliveGreen}{rgb}{0.33,0.42,0.18}
\definecolor{DarkOrange1}{rgb}{1.00,0.50,0.00}
\definecolor{DarkOrange2}{rgb}{0.93,0.46,0.00}
\definecolor{DarkOrange3}{rgb}{0.80,0.40,0.00}
\definecolor{DarkOrange4}{rgb}{0.55,0.27,0.00}
\definecolor{DarkOrange}{rgb}{1.00,0.55,0.00}
\definecolor{DarkOrchid1}{rgb}{0.75,0.24,1.00}
\definecolor{DarkOrchid2}{rgb}{0.70,0.23,0.93}
\definecolor{DarkOrchid3}{rgb}{0.60,0.20,0.80}
\definecolor{DarkOrchid4}{rgb}{0.41,0.13,0.55}
\definecolor{DarkOrchid}{rgb}{0.60,0.20,0.80}
\definecolor{DarkRed}{rgb}{0.55,0.00,0.00}
\definecolor{DarkSalmon}{rgb}{0.91,0.59,0.48}
\definecolor{DarkSeaGreen1}{rgb}{0.76,1.00,0.76}
\definecolor{DarkSeaGreen2}{rgb}{0.71,0.93,0.71}
\definecolor{DarkSeaGreen3}{rgb}{0.61,0.80,0.61}
\definecolor{DarkSeaGreen4}{rgb}{0.41,0.55,0.41}
\definecolor{DarkSeaGreen}{rgb}{0.56,0.74,0.56}
\definecolor{DarkSlateBlue}{rgb}{0.28,0.24,0.55}
\definecolor{DarkSlateGray1}{rgb}{0.59,1.00,1.00}
\definecolor{DarkSlateGray2}{rgb}{0.55,0.93,0.93}
\definecolor{DarkSlateGray3}{rgb}{0.47,0.80,0.80}
\definecolor{DarkSlateGray4}{rgb}{0.32,0.55,0.55}
\definecolor{DarkSlateGray}{rgb}{0.18,0.31,0.31}
\definecolor{DarkSlateGrey}{rgb}{0.18,0.31,0.31}
\definecolor{DarkTurquoise}{rgb}{0.00,0.81,0.82}
\definecolor{DarkViolet}{rgb}{0.58,0.00,0.83}
\definecolor{DeepPink1}{rgb}{1.00,0.08,0.58}
\definecolor{DeepPink2}{rgb}{0.93,0.07,0.54}
\definecolor{DeepPink3}{rgb}{0.80,0.06,0.46}
\definecolor{DeepPink4}{rgb}{0.55,0.04,0.31}
\definecolor{DeepPink}{rgb}{1.00,0.08,0.58}
\definecolor{DeepSkyBlue1}{rgb}{0.00,0.75,1.00}
\definecolor{DeepSkyBlue2}{rgb}{0.00,0.70,0.93}
\definecolor{DeepSkyBlue3}{rgb}{0.00,0.60,0.80}
\definecolor{DeepSkyBlue4}{rgb}{0.00,0.41,0.55}
\definecolor{DeepSkyBlue}{rgb}{0.00,0.75,1.00}
\definecolor{DimGray}{rgb}{0.41,0.41,0.41}
\definecolor{DimGrey}{rgb}{0.41,0.41,0.41}
\definecolor{DodgerBlue1}{rgb}{0.12,0.56,1.00}
\definecolor{DodgerBlue2}{rgb}{0.11,0.53,0.93}
\definecolor{DodgerBlue3}{rgb}{0.09,0.45,0.80}
\definecolor{DodgerBlue4}{rgb}{0.06,0.31,0.55}
\definecolor{DodgerBlue}{rgb}{0.12,0.56,1.00}
\definecolor{FloralWhite}{rgb}{1.00,0.98,0.94}
\definecolor{ForestGreen}{rgb}{0.13,0.55,0.13}
\definecolor{GhostWhite}{rgb}{0.97,0.97,1.00}
\definecolor{GreenYellow}{rgb}{0.68,1.00,0.18}
\definecolor{HotPink1}{rgb}{1.00,0.43,0.71}
\definecolor{HotPink2}{rgb}{0.93,0.42,0.65}
\definecolor{HotPink3}{rgb}{0.80,0.38,0.56}
\definecolor{HotPink4}{rgb}{0.55,0.23,0.38}
\definecolor{HotPink}{rgb}{1.00,0.41,0.71}
\definecolor{IndianRed1}{rgb}{1.00,0.42,0.42}
\definecolor{IndianRed2}{rgb}{0.93,0.39,0.39}
\definecolor{IndianRed3}{rgb}{0.80,0.33,0.33}
\definecolor{IndianRed4}{rgb}{0.55,0.23,0.23}
\definecolor{IndianRed}{rgb}{0.80,0.36,0.36}
\definecolor{LavenderBlush1}{rgb}{1.00,0.94,0.96}
\definecolor{LavenderBlush2}{rgb}{0.93,0.88,0.90}
\definecolor{LavenderBlush3}{rgb}{0.80,0.76,0.77}
\definecolor{LavenderBlush4}{rgb}{0.55,0.51,0.53}
\definecolor{LavenderBlush}{rgb}{1.00,0.94,0.96}
\definecolor{LawnGreen}{rgb}{0.49,0.99,0.00}
\definecolor{LemonChiffon1}{rgb}{1.00,0.98,0.80}
\definecolor{LemonChiffon2}{rgb}{0.93,0.91,0.75}
\definecolor{LemonChiffon3}{rgb}{0.80,0.79,0.65}
\definecolor{LemonChiffon4}{rgb}{0.55,0.54,0.44}
\definecolor{LemonChiffon}{rgb}{1.00,0.98,0.80}
\definecolor{LightBlue1}{rgb}{0.75,0.94,1.00}
\definecolor{LightBlue2}{rgb}{0.70,0.87,0.93}
\definecolor{LightBlue3}{rgb}{0.60,0.75,0.80}
\definecolor{LightBlue4}{rgb}{0.41,0.51,0.55}
\definecolor{LightBlue}{rgb}{0.68,0.85,0.90}
\definecolor{LightCoral}{rgb}{0.94,0.50,0.50}
\definecolor{LightCyan1}{rgb}{0.88,1.00,1.00}
\definecolor{LightCyan2}{rgb}{0.82,0.93,0.93}
\definecolor{LightCyan3}{rgb}{0.71,0.80,0.80}
\definecolor{LightCyan4}{rgb}{0.48,0.55,0.55}
\definecolor{LightCyan}{rgb}{0.88,1.00,1.00}
\definecolor{LightGoldenrod1}{rgb}{1.00,0.93,0.55}
\definecolor{LightGoldenrod2}{rgb}{0.93,0.86,0.51}
\definecolor{LightGoldenrod3}{rgb}{0.80,0.75,0.44}
\definecolor{LightGoldenrod4}{rgb}{0.55,0.51,0.30}
\definecolor{LightGoldenrodYellow}{rgb}{0.98,0.98,0.82}
\definecolor{LightGoldenrod}{rgb}{0.93,0.87,0.51}
\definecolor{LightGray}{rgb}{0.83,0.83,0.83}
\definecolor{LightGreen}{rgb}{0.56,0.93,0.56}
\definecolor{LightGrey}{rgb}{0.83,0.83,0.83}
\definecolor{LightPink1}{rgb}{1.00,0.68,0.73}
\definecolor{LightPink2}{rgb}{0.93,0.64,0.68}
\definecolor{LightPink3}{rgb}{0.80,0.55,0.58}
\definecolor{LightPink4}{rgb}{0.55,0.37,0.40}
\definecolor{LightPink}{rgb}{1.00,0.71,0.76}
\definecolor{LightSalmon1}{rgb}{1.00,0.63,0.48}
\definecolor{LightSalmon2}{rgb}{0.93,0.58,0.45}
\definecolor{LightSalmon3}{rgb}{0.80,0.51,0.38}
\definecolor{LightSalmon4}{rgb}{0.55,0.34,0.26}
\definecolor{LightSalmon}{rgb}{1.00,0.63,0.48}
\definecolor{LightSeaGreen}{rgb}{0.13,0.70,0.67}
\definecolor{LightSkyBlue1}{rgb}{0.69,0.89,1.00}
\definecolor{LightSkyBlue2}{rgb}{0.64,0.83,0.93}
\definecolor{LightSkyBlue3}{rgb}{0.55,0.71,0.80}
\definecolor{LightSkyBlue4}{rgb}{0.38,0.48,0.55}
\definecolor{LightSkyBlue}{rgb}{0.53,0.81,0.98}
\definecolor{LightSlateBlue}{rgb}{0.52,0.44,1.00}
\definecolor{LightSlateGray}{rgb}{0.47,0.53,0.60}
\definecolor{LightSlateGrey}{rgb}{0.47,0.53,0.60}
\definecolor{LightSteelBlue1}{rgb}{0.79,0.88,1.00}
\definecolor{LightSteelBlue2}{rgb}{0.74,0.82,0.93}
\definecolor{LightSteelBlue3}{rgb}{0.64,0.71,0.80}
\definecolor{LightSteelBlue4}{rgb}{0.43,0.48,0.55}
\definecolor{LightSteelBlue}{rgb}{0.69,0.77,0.87}
\definecolor{LightYellow1}{rgb}{1.00,1.00,0.88}
\definecolor{LightYellow2}{rgb}{0.93,0.93,0.82}
\definecolor{LightYellow3}{rgb}{0.80,0.80,0.71}
\definecolor{LightYellow4}{rgb}{0.55,0.55,0.48}
\definecolor{LightYellow}{rgb}{1.00,1.00,0.88}
\definecolor{LimeGreen}{rgb}{0.20,0.80,0.20}
\definecolor{MediumAquamarine}{rgb}{0.40,0.80,0.67}
\definecolor{MediumBlue}{rgb}{0.00,0.00,0.80}
\definecolor{MediumOrchid1}{rgb}{0.88,0.40,1.00}
\definecolor{MediumOrchid2}{rgb}{0.82,0.37,0.93}
\definecolor{MediumOrchid3}{rgb}{0.71,0.32,0.80}
\definecolor{MediumOrchid4}{rgb}{0.48,0.22,0.55}
\definecolor{MediumOrchid}{rgb}{0.73,0.33,0.83}
\definecolor{MediumPurple1}{rgb}{0.67,0.51,1.00}
\definecolor{MediumPurple2}{rgb}{0.62,0.47,0.93}
\definecolor{MediumPurple3}{rgb}{0.54,0.41,0.80}
\definecolor{MediumPurple4}{rgb}{0.36,0.28,0.55}
\definecolor{MediumPurple}{rgb}{0.58,0.44,0.86}
\definecolor{MediumSeaGreen}{rgb}{0.24,0.70,0.44}
\definecolor{MediumSlateBlue}{rgb}{0.48,0.41,0.93}
\definecolor{MediumSpringGreen}{rgb}{0.00,0.98,0.60}
\definecolor{MediumTurquoise}{rgb}{0.28,0.82,0.80}
\definecolor{MediumVioletRed}{rgb}{0.78,0.08,0.52}
\definecolor{MidnightBlue}{rgb}{0.10,0.10,0.44}
\definecolor{MintCream}{rgb}{0.96,1.00,0.98}
\definecolor{MistyRose1}{rgb}{1.00,0.89,0.88}
\definecolor{MistyRose2}{rgb}{0.93,0.84,0.82}
\definecolor{MistyRose3}{rgb}{0.80,0.72,0.71}
\definecolor{MistyRose4}{rgb}{0.55,0.49,0.48}
\definecolor{MistyRose}{rgb}{1.00,0.89,0.88}
\definecolor{NavajoWhite1}{rgb}{1.00,0.87,0.68}
\definecolor{NavajoWhite2}{rgb}{0.93,0.81,0.63}
\definecolor{NavajoWhite3}{rgb}{0.80,0.70,0.55}
\definecolor{NavajoWhite4}{rgb}{0.55,0.47,0.37}
\definecolor{NavajoWhite}{rgb}{1.00,0.87,0.68}
\definecolor{NavyBlue}{rgb}{0.00,0.00,0.50}
\definecolor{OldLace}{rgb}{0.99,0.96,0.90}
\definecolor{OliveDrab1}{rgb}{0.75,1.00,0.24}
\definecolor{OliveDrab2}{rgb}{0.70,0.93,0.23}
\definecolor{OliveDrab3}{rgb}{0.60,0.80,0.20}
\definecolor{OliveDrab4}{rgb}{0.41,0.55,0.13}
\definecolor{OliveDrab}{rgb}{0.42,0.56,0.14}
\definecolor{OrangeRed1}{rgb}{1.00,0.27,0.00}
\definecolor{OrangeRed2}{rgb}{0.93,0.25,0.00}
\definecolor{OrangeRed3}{rgb}{0.80,0.22,0.00}
\definecolor{OrangeRed4}{rgb}{0.55,0.15,0.00}
\definecolor{OrangeRed}{rgb}{1.00,0.27,0.00}
\definecolor{PaleGoldenrod}{rgb}{0.93,0.91,0.67}
\definecolor{PaleGreen1}{rgb}{0.60,1.00,0.60}
\definecolor{PaleGreen2}{rgb}{0.56,0.93,0.56}
\definecolor{PaleGreen3}{rgb}{0.49,0.80,0.49}
\definecolor{PaleGreen4}{rgb}{0.33,0.55,0.33}
\definecolor{PaleGreen}{rgb}{0.60,0.98,0.60}
\definecolor{PaleTurquoise1}{rgb}{0.73,1.00,1.00}
\definecolor{PaleTurquoise2}{rgb}{0.68,0.93,0.93}
\definecolor{PaleTurquoise3}{rgb}{0.59,0.80,0.80}
\definecolor{PaleTurquoise4}{rgb}{0.40,0.55,0.55}
\definecolor{PaleTurquoise}{rgb}{0.69,0.93,0.93}
\definecolor{PaleVioletRed1}{rgb}{1.00,0.51,0.67}
\definecolor{PaleVioletRed2}{rgb}{0.93,0.47,0.62}
\definecolor{PaleVioletRed3}{rgb}{0.80,0.41,0.54}
\definecolor{PaleVioletRed4}{rgb}{0.55,0.28,0.36}
\definecolor{PaleVioletRed}{rgb}{0.86,0.44,0.58}
\definecolor{PapayaWhip}{rgb}{1.00,0.94,0.84}
\definecolor{PeachPuff1}{rgb}{1.00,0.85,0.73}
\definecolor{PeachPuff2}{rgb}{0.93,0.80,0.68}
\definecolor{PeachPuff3}{rgb}{0.80,0.69,0.58}
\definecolor{PeachPuff4}{rgb}{0.55,0.47,0.40}
\definecolor{PeachPuff}{rgb}{1.00,0.85,0.73}
\definecolor{PowderBlue}{rgb}{0.69,0.88,0.90}
\definecolor{RosyBrown1}{rgb}{1.00,0.76,0.76}
\definecolor{RosyBrown2}{rgb}{0.93,0.71,0.71}
\definecolor{RosyBrown3}{rgb}{0.80,0.61,0.61}
\definecolor{RosyBrown4}{rgb}{0.55,0.41,0.41}
\definecolor{RosyBrown}{rgb}{0.74,0.56,0.56}
\definecolor{RoyalBlue1}{rgb}{0.28,0.46,1.00}
\definecolor{RoyalBlue2}{rgb}{0.26,0.43,0.93}
\definecolor{RoyalBlue3}{rgb}{0.23,0.37,0.80}
\definecolor{RoyalBlue4}{rgb}{0.15,0.25,0.55}
\definecolor{RoyalBlue}{rgb}{0.25,0.41,0.88}
\definecolor{SaddleBrown}{rgb}{0.55,0.27,0.07}
\definecolor{SandyBrown}{rgb}{0.96,0.64,0.38}
\definecolor{SeaGreen1}{rgb}{0.33,1.00,0.62}
\definecolor{SeaGreen2}{rgb}{0.31,0.93,0.58}
\definecolor{SeaGreen3}{rgb}{0.26,0.80,0.50}
\definecolor{SeaGreen4}{rgb}{0.18,0.55,0.34}
\definecolor{SeaGreen}{rgb}{0.18,0.55,0.34}
\definecolor{SkyBlue1}{rgb}{0.53,0.81,1.00}
\definecolor{SkyBlue2}{rgb}{0.49,0.75,0.93}
\definecolor{SkyBlue3}{rgb}{0.42,0.65,0.80}
\definecolor{SkyBlue4}{rgb}{0.29,0.44,0.55}
\definecolor{SkyBlue}{rgb}{0.53,0.81,0.92}
\definecolor{SlateBlue1}{rgb}{0.51,0.44,1.00}
\definecolor{SlateBlue2}{rgb}{0.48,0.40,0.93}
\definecolor{SlateBlue3}{rgb}{0.41,0.35,0.80}
\definecolor{SlateBlue4}{rgb}{0.28,0.24,0.55}
\definecolor{SlateBlue}{rgb}{0.42,0.35,0.80}
\definecolor{SlateGray1}{rgb}{0.78,0.89,1.00}
\definecolor{SlateGray2}{rgb}{0.73,0.83,0.93}
\definecolor{SlateGray3}{rgb}{0.62,0.71,0.80}
\definecolor{SlateGray4}{rgb}{0.42,0.48,0.55}
\definecolor{SlateGray}{rgb}{0.44,0.50,0.56}
\definecolor{SlateGrey}{rgb}{0.44,0.50,0.56}
\definecolor{SpringGreen1}{rgb}{0.00,1.00,0.50}
\definecolor{SpringGreen2}{rgb}{0.00,0.93,0.46}
\definecolor{SpringGreen3}{rgb}{0.00,0.80,0.40}
\definecolor{SpringGreen4}{rgb}{0.00,0.55,0.27}
\definecolor{SpringGreen}{rgb}{0.00,1.00,0.50}
\definecolor{SteelBlue1}{rgb}{0.39,0.72,1.00}
\definecolor{SteelBlue2}{rgb}{0.36,0.67,0.93}
\definecolor{SteelBlue3}{rgb}{0.31,0.58,0.80}
\definecolor{SteelBlue4}{rgb}{0.21,0.39,0.55}
\definecolor{SteelBlue}{rgb}{0.27,0.51,0.71}
\definecolor{VioletRed1}{rgb}{1.00,0.24,0.59}
\definecolor{VioletRed2}{rgb}{0.93,0.23,0.55}
\definecolor{VioletRed3}{rgb}{0.80,0.20,0.47}
\definecolor{VioletRed4}{rgb}{0.55,0.13,0.32}
\definecolor{VioletRed}{rgb}{0.82,0.13,0.56}
\definecolor{WhiteSmoke}{rgb}{0.96,0.96,0.96}
\definecolor{YellowGreen}{rgb}{0.60,0.80,0.20}
\definecolor{aliceblue}{rgb}{0.94,0.97,1.00}
\definecolor{antiquewhite}{rgb}{0.98,0.92,0.84}
\definecolor{aquamarine1}{rgb}{0.50,1.00,0.83}
\definecolor{aquamarine2}{rgb}{0.46,0.93,0.78}
\definecolor{aquamarine3}{rgb}{0.40,0.80,0.67}
\definecolor{aquamarine4}{rgb}{0.27,0.55,0.45}
\definecolor{aquamarine}{rgb}{0.50,1.00,0.83}
\definecolor{azure1}{rgb}{0.94,1.00,1.00}
\definecolor{azure2}{rgb}{0.88,0.93,0.93}
\definecolor{azure3}{rgb}{0.76,0.80,0.80}
\definecolor{azure4}{rgb}{0.51,0.55,0.55}
\definecolor{azure}{rgb}{0.94,1.00,1.00}
\definecolor{beige}{rgb}{0.96,0.96,0.86}
\definecolor{bisque1}{rgb}{1.00,0.89,0.77}
\definecolor{bisque2}{rgb}{0.93,0.84,0.72}
\definecolor{bisque3}{rgb}{0.80,0.72,0.62}
\definecolor{bisque4}{rgb}{0.55,0.49,0.42}
\definecolor{bisque}{rgb}{1.00,0.89,0.77}
\definecolor{black}{rgb}{0.00,0.00,0.00}
\definecolor{blanchedalmond}{rgb}{1.00,0.92,0.80}
\definecolor{blue1}{rgb}{0.00,0.00,1.00}
\definecolor{blue2}{rgb}{0.00,0.00,0.93}
\definecolor{blue3}{rgb}{0.00,0.00,0.80}
\definecolor{blue4}{rgb}{0.00,0.00,0.55}
\definecolor{blueviolet}{rgb}{0.54,0.17,0.89}
\definecolor{blue}{rgb}{0.00,0.00,1.00}
\definecolor{brown1}{rgb}{1.00,0.25,0.25}
\definecolor{brown2}{rgb}{0.93,0.23,0.23}
\definecolor{brown3}{rgb}{0.80,0.20,0.20}
\definecolor{brown4}{rgb}{0.55,0.14,0.14}
\definecolor{brown}{rgb}{0.65,0.16,0.16}
\definecolor{burlywood1}{rgb}{1.00,0.83,0.61}
\definecolor{burlywood2}{rgb}{0.93,0.77,0.57}
\definecolor{burlywood3}{rgb}{0.80,0.67,0.49}
\definecolor{burlywood4}{rgb}{0.55,0.45,0.33}
\definecolor{burlywood}{rgb}{0.87,0.72,0.53}
\definecolor{cadetblue}{rgb}{0.37,0.62,0.63}
\definecolor{chartreuse1}{rgb}{0.50,1.00,0.00}
\definecolor{chartreuse2}{rgb}{0.46,0.93,0.00}
\definecolor{chartreuse3}{rgb}{0.40,0.80,0.00}
\definecolor{chartreuse4}{rgb}{0.27,0.55,0.00}
\definecolor{chartreuse}{rgb}{0.50,1.00,0.00}
\definecolor{chocolate1}{rgb}{1.00,0.50,0.14}
\definecolor{chocolate2}{rgb}{0.93,0.46,0.13}
\definecolor{chocolate3}{rgb}{0.80,0.40,0.11}
\definecolor{chocolate4}{rgb}{0.55,0.27,0.07}
\definecolor{chocolate}{rgb}{0.82,0.41,0.12}
\definecolor{coral1}{rgb}{1.00,0.45,0.34}
\definecolor{coral2}{rgb}{0.93,0.42,0.31}
\definecolor{coral3}{rgb}{0.80,0.36,0.27}
\definecolor{coral4}{rgb}{0.55,0.24,0.18}
\definecolor{coral}{rgb}{1.00,0.50,0.31}
\definecolor{cornflowerblue}{rgb}{0.39,0.58,0.93}
\definecolor{cornsilk1}{rgb}{1.00,0.97,0.86}
\definecolor{cornsilk2}{rgb}{0.93,0.91,0.80}
\definecolor{cornsilk3}{rgb}{0.80,0.78,0.69}
\definecolor{cornsilk4}{rgb}{0.55,0.53,0.47}
\definecolor{cornsilk}{rgb}{1.00,0.97,0.86}
\definecolor{cyan1}{rgb}{0.00,1.00,1.00}
\definecolor{cyan2}{rgb}{0.00,0.93,0.93}
\definecolor{cyan3}{rgb}{0.00,0.80,0.80}
\definecolor{cyan4}{rgb}{0.00,0.55,0.55}
\definecolor{cyan}{rgb}{0.00,1.00,1.00}
\definecolor{darkblue}{rgb}{0.00,0.00,0.55}
\definecolor{darkcyan}{rgb}{0.00,0.55,0.55}
\definecolor{darkgoldenrod}{rgb}{0.72,0.53,0.04}
\definecolor{darkgray}{rgb}{0.66,0.66,0.66}
\definecolor{darkgreen}{rgb}{0.00,0.39,0.00}
\definecolor{darkgrey}{rgb}{0.66,0.66,0.66}
\definecolor{darkkhaki}{rgb}{0.74,0.72,0.42}
\definecolor{darkmagenta}{rgb}{0.55,0.00,0.55}
\definecolor{darkolive}{rgb}{0.33,0.42,0.18}
\definecolor{darkorange}{rgb}{1.00,0.55,0.00}
\definecolor{darkorchid}{rgb}{0.60,0.20,0.80}
\definecolor{darkred}{rgb}{0.55,0.00,0.00}
\definecolor{darksalmon}{rgb}{0.91,0.59,0.48}
\definecolor{darksea}{rgb}{0.56,0.74,0.56}
\definecolor{darkslate}{rgb}{0.18,0.31,0.31}
\definecolor{darkslate}{rgb}{0.18,0.31,0.31}
\definecolor{darkslate}{rgb}{0.28,0.24,0.55}
\definecolor{darkturquoise}{rgb}{0.00,0.81,0.82}
\definecolor{darkviolet}{rgb}{0.58,0.00,0.83}
\definecolor{deeppink}{rgb}{1.00,0.08,0.58}
\definecolor{deepsky}{rgb}{0.00,0.75,1.00}
\definecolor{dimgray}{rgb}{0.41,0.41,0.41}
\definecolor{dimgrey}{rgb}{0.41,0.41,0.41}
\definecolor{dodgerblue}{rgb}{0.12,0.56,1.00}
\definecolor{firebrick1}{rgb}{1.00,0.19,0.19}
\definecolor{firebrick2}{rgb}{0.93,0.17,0.17}
\definecolor{firebrick3}{rgb}{0.80,0.15,0.15}
\definecolor{firebrick4}{rgb}{0.55,0.10,0.10}
\definecolor{firebrick}{rgb}{0.70,0.13,0.13}
\definecolor{floralwhite}{rgb}{1.00,0.98,0.94}
\definecolor{forestgreen}{rgb}{0.13,0.55,0.13}
\definecolor{gainsboro}{rgb}{0.86,0.86,0.86}
\definecolor{ghostwhite}{rgb}{0.97,0.97,1.00}
\definecolor{gold1}{rgb}{1.00,0.84,0.00}
\definecolor{gold2}{rgb}{0.93,0.79,0.00}
\definecolor{gold3}{rgb}{0.80,0.68,0.00}
\definecolor{gold4}{rgb}{0.55,0.46,0.00}
\definecolor{goldenrod1}{rgb}{1.00,0.76,0.15}
\definecolor{goldenrod2}{rgb}{0.93,0.71,0.13}
\definecolor{goldenrod3}{rgb}{0.80,0.61,0.11}
\definecolor{goldenrod4}{rgb}{0.55,0.41,0.08}
\definecolor{goldenrod}{rgb}{0.85,0.65,0.13}
\definecolor{gold}{rgb}{1.00,0.84,0.00}
\definecolor{gray0}{rgb}{0.00,0.00,0.00}
\definecolor{gray100}{rgb}{1.00,1.00,1.00}
\definecolor{gray10}{rgb}{0.10,0.10,0.10}
\definecolor{gray11}{rgb}{0.11,0.11,0.11}
\definecolor{gray12}{rgb}{0.12,0.12,0.12}
\definecolor{gray13}{rgb}{0.13,0.13,0.13}
\definecolor{gray14}{rgb}{0.14,0.14,0.14}
\definecolor{gray15}{rgb}{0.15,0.15,0.15}
\definecolor{gray16}{rgb}{0.16,0.16,0.16}
\definecolor{gray17}{rgb}{0.17,0.17,0.17}
\definecolor{gray18}{rgb}{0.18,0.18,0.18}
\definecolor{gray19}{rgb}{0.19,0.19,0.19}
\definecolor{gray1}{rgb}{0.01,0.01,0.01}
\definecolor{gray20}{rgb}{0.20,0.20,0.20}
\definecolor{gray21}{rgb}{0.21,0.21,0.21}
\definecolor{gray22}{rgb}{0.22,0.22,0.22}
\definecolor{gray23}{rgb}{0.23,0.23,0.23}
\definecolor{gray24}{rgb}{0.24,0.24,0.24}
\definecolor{gray25}{rgb}{0.25,0.25,0.25}
\definecolor{gray26}{rgb}{0.26,0.26,0.26}
\definecolor{gray27}{rgb}{0.27,0.27,0.27}
\definecolor{gray28}{rgb}{0.28,0.28,0.28}
\definecolor{gray29}{rgb}{0.29,0.29,0.29}
\definecolor{gray2}{rgb}{0.02,0.02,0.02}
\definecolor{gray30}{rgb}{0.30,0.30,0.30}
\definecolor{gray31}{rgb}{0.31,0.31,0.31}
\definecolor{gray32}{rgb}{0.32,0.32,0.32}
\definecolor{gray33}{rgb}{0.33,0.33,0.33}
\definecolor{gray34}{rgb}{0.34,0.34,0.34}
\definecolor{gray35}{rgb}{0.35,0.35,0.35}
\definecolor{gray36}{rgb}{0.36,0.36,0.36}
\definecolor{gray37}{rgb}{0.37,0.37,0.37}
\definecolor{gray38}{rgb}{0.38,0.38,0.38}
\definecolor{gray39}{rgb}{0.39,0.39,0.39}
\definecolor{gray3}{rgb}{0.03,0.03,0.03}
\definecolor{gray40}{rgb}{0.40,0.40,0.40}
\definecolor{gray41}{rgb}{0.41,0.41,0.41}
\definecolor{gray42}{rgb}{0.42,0.42,0.42}
\definecolor{gray43}{rgb}{0.43,0.43,0.43}
\definecolor{gray44}{rgb}{0.44,0.44,0.44}
\definecolor{gray45}{rgb}{0.45,0.45,0.45}
\definecolor{gray46}{rgb}{0.46,0.46,0.46}
\definecolor{gray47}{rgb}{0.47,0.47,0.47}
\definecolor{gray48}{rgb}{0.48,0.48,0.48}
\definecolor{gray49}{rgb}{0.49,0.49,0.49}
\definecolor{gray4}{rgb}{0.04,0.04,0.04}
\definecolor{gray50}{rgb}{0.50,0.50,0.50}
\definecolor{gray51}{rgb}{0.51,0.51,0.51}
\definecolor{gray52}{rgb}{0.52,0.52,0.52}
\definecolor{gray53}{rgb}{0.53,0.53,0.53}
\definecolor{gray54}{rgb}{0.54,0.54,0.54}
\definecolor{gray55}{rgb}{0.55,0.55,0.55}
\definecolor{gray56}{rgb}{0.56,0.56,0.56}
\definecolor{gray57}{rgb}{0.57,0.57,0.57}
\definecolor{gray58}{rgb}{0.58,0.58,0.58}
\definecolor{gray59}{rgb}{0.59,0.59,0.59}
\definecolor{gray5}{rgb}{0.05,0.05,0.05}
\definecolor{gray60}{rgb}{0.60,0.60,0.60}
\definecolor{gray61}{rgb}{0.61,0.61,0.61}
\definecolor{gray62}{rgb}{0.62,0.62,0.62}
\definecolor{gray63}{rgb}{0.63,0.63,0.63}
\definecolor{gray64}{rgb}{0.64,0.64,0.64}
\definecolor{gray65}{rgb}{0.65,0.65,0.65}
\definecolor{gray66}{rgb}{0.66,0.66,0.66}
\definecolor{gray67}{rgb}{0.67,0.67,0.67}
\definecolor{gray68}{rgb}{0.68,0.68,0.68}
\definecolor{gray69}{rgb}{0.69,0.69,0.69}
\definecolor{gray6}{rgb}{0.06,0.06,0.06}
\definecolor{gray70}{rgb}{0.70,0.70,0.70}
\definecolor{gray71}{rgb}{0.71,0.71,0.71}
\definecolor{gray72}{rgb}{0.72,0.72,0.72}
\definecolor{gray73}{rgb}{0.73,0.73,0.73}
\definecolor{gray74}{rgb}{0.74,0.74,0.74}
\definecolor{gray75}{rgb}{0.75,0.75,0.75}
\definecolor{gray76}{rgb}{0.76,0.76,0.76}
\definecolor{gray77}{rgb}{0.77,0.77,0.77}
\definecolor{gray78}{rgb}{0.78,0.78,0.78}
\definecolor{gray79}{rgb}{0.79,0.79,0.79}
\definecolor{gray7}{rgb}{0.07,0.07,0.07}
\definecolor{gray80}{rgb}{0.80,0.80,0.80}
\definecolor{gray81}{rgb}{0.81,0.81,0.81}
\definecolor{gray82}{rgb}{0.82,0.82,0.82}
\definecolor{gray83}{rgb}{0.83,0.83,0.83}
\definecolor{gray84}{rgb}{0.84,0.84,0.84}
\definecolor{gray85}{rgb}{0.85,0.85,0.85}
\definecolor{gray86}{rgb}{0.86,0.86,0.86}
\definecolor{gray87}{rgb}{0.87,0.87,0.87}
\definecolor{gray88}{rgb}{0.88,0.88,0.88}
\definecolor{gray89}{rgb}{0.89,0.89,0.89}
\definecolor{gray8}{rgb}{0.08,0.08,0.08}
\definecolor{gray90}{rgb}{0.90,0.90,0.90}
\definecolor{gray91}{rgb}{0.91,0.91,0.91}
\definecolor{gray92}{rgb}{0.92,0.92,0.92}
\definecolor{gray93}{rgb}{0.93,0.93,0.93}
\definecolor{gray94}{rgb}{0.94,0.94,0.94}
\definecolor{gray95}{rgb}{0.95,0.95,0.95}
\definecolor{gray96}{rgb}{0.96,0.96,0.96}
\definecolor{gray97}{rgb}{0.97,0.97,0.97}
\definecolor{gray98}{rgb}{0.98,0.98,0.98}
\definecolor{gray99}{rgb}{0.99,0.99,0.99}
\definecolor{gray9}{rgb}{0.09,0.09,0.09}
\definecolor{gray}{rgb}{0.75,0.75,0.75}
\definecolor{green1}{rgb}{0.00,1.00,0.00}
\definecolor{green2}{rgb}{0.00,0.93,0.00}
\definecolor{green3}{rgb}{0.00,0.80,0.00}
\definecolor{green4}{rgb}{0.00,0.55,0.00}
\definecolor{greenyellow}{rgb}{0.68,1.00,0.18}
\definecolor{green}{rgb}{0.00,1.00,0.00}
\definecolor{grey0}{rgb}{0.00,0.00,0.00}
\definecolor{grey100}{rgb}{1.00,1.00,1.00}
\definecolor{grey10}{rgb}{0.10,0.10,0.10}
\definecolor{grey11}{rgb}{0.11,0.11,0.11}
\definecolor{grey12}{rgb}{0.12,0.12,0.12}
\definecolor{grey13}{rgb}{0.13,0.13,0.13}
\definecolor{grey14}{rgb}{0.14,0.14,0.14}
\definecolor{grey15}{rgb}{0.15,0.15,0.15}
\definecolor{grey16}{rgb}{0.16,0.16,0.16}
\definecolor{grey17}{rgb}{0.17,0.17,0.17}
\definecolor{grey18}{rgb}{0.18,0.18,0.18}
\definecolor{grey19}{rgb}{0.19,0.19,0.19}
\definecolor{grey1}{rgb}{0.01,0.01,0.01}
\definecolor{grey20}{rgb}{0.20,0.20,0.20}
\definecolor{grey21}{rgb}{0.21,0.21,0.21}
\definecolor{grey22}{rgb}{0.22,0.22,0.22}
\definecolor{grey23}{rgb}{0.23,0.23,0.23}
\definecolor{grey24}{rgb}{0.24,0.24,0.24}
\definecolor{grey25}{rgb}{0.25,0.25,0.25}
\definecolor{grey26}{rgb}{0.26,0.26,0.26}
\definecolor{grey27}{rgb}{0.27,0.27,0.27}
\definecolor{grey28}{rgb}{0.28,0.28,0.28}
\definecolor{grey29}{rgb}{0.29,0.29,0.29}
\definecolor{grey2}{rgb}{0.02,0.02,0.02}
\definecolor{grey30}{rgb}{0.30,0.30,0.30}
\definecolor{grey31}{rgb}{0.31,0.31,0.31}
\definecolor{grey32}{rgb}{0.32,0.32,0.32}
\definecolor{grey33}{rgb}{0.33,0.33,0.33}
\definecolor{grey34}{rgb}{0.34,0.34,0.34}
\definecolor{grey35}{rgb}{0.35,0.35,0.35}
\definecolor{grey36}{rgb}{0.36,0.36,0.36}
\definecolor{grey37}{rgb}{0.37,0.37,0.37}
\definecolor{grey38}{rgb}{0.38,0.38,0.38}
\definecolor{grey39}{rgb}{0.39,0.39,0.39}
\definecolor{grey3}{rgb}{0.03,0.03,0.03}
\definecolor{grey40}{rgb}{0.40,0.40,0.40}
\definecolor{grey41}{rgb}{0.41,0.41,0.41}
\definecolor{grey42}{rgb}{0.42,0.42,0.42}
\definecolor{grey43}{rgb}{0.43,0.43,0.43}
\definecolor{grey44}{rgb}{0.44,0.44,0.44}
\definecolor{grey45}{rgb}{0.45,0.45,0.45}
\definecolor{grey46}{rgb}{0.46,0.46,0.46}
\definecolor{grey47}{rgb}{0.47,0.47,0.47}
\definecolor{grey48}{rgb}{0.48,0.48,0.48}
\definecolor{grey49}{rgb}{0.49,0.49,0.49}
\definecolor{grey4}{rgb}{0.04,0.04,0.04}
\definecolor{grey50}{rgb}{0.50,0.50,0.50}
\definecolor{grey51}{rgb}{0.51,0.51,0.51}
\definecolor{grey52}{rgb}{0.52,0.52,0.52}
\definecolor{grey53}{rgb}{0.53,0.53,0.53}
\definecolor{grey54}{rgb}{0.54,0.54,0.54}
\definecolor{grey55}{rgb}{0.55,0.55,0.55}
\definecolor{grey56}{rgb}{0.56,0.56,0.56}
\definecolor{grey57}{rgb}{0.57,0.57,0.57}
\definecolor{grey58}{rgb}{0.58,0.58,0.58}
\definecolor{grey59}{rgb}{0.59,0.59,0.59}
\definecolor{grey5}{rgb}{0.05,0.05,0.05}
\definecolor{grey60}{rgb}{0.60,0.60,0.60}
\definecolor{grey61}{rgb}{0.61,0.61,0.61}
\definecolor{grey62}{rgb}{0.62,0.62,0.62}
\definecolor{grey63}{rgb}{0.63,0.63,0.63}
\definecolor{grey64}{rgb}{0.64,0.64,0.64}
\definecolor{grey65}{rgb}{0.65,0.65,0.65}
\definecolor{grey66}{rgb}{0.66,0.66,0.66}
\definecolor{grey67}{rgb}{0.67,0.67,0.67}
\definecolor{grey68}{rgb}{0.68,0.68,0.68}
\definecolor{grey69}{rgb}{0.69,0.69,0.69}
\definecolor{grey6}{rgb}{0.06,0.06,0.06}
\definecolor{grey70}{rgb}{0.70,0.70,0.70}
\definecolor{grey71}{rgb}{0.71,0.71,0.71}
\definecolor{grey72}{rgb}{0.72,0.72,0.72}
\definecolor{grey73}{rgb}{0.73,0.73,0.73}
\definecolor{grey74}{rgb}{0.74,0.74,0.74}
\definecolor{grey75}{rgb}{0.75,0.75,0.75}
\definecolor{grey76}{rgb}{0.76,0.76,0.76}
\definecolor{grey77}{rgb}{0.77,0.77,0.77}
\definecolor{grey78}{rgb}{0.78,0.78,0.78}
\definecolor{grey79}{rgb}{0.79,0.79,0.79}
\definecolor{grey7}{rgb}{0.07,0.07,0.07}
\definecolor{grey80}{rgb}{0.80,0.80,0.80}
\definecolor{grey81}{rgb}{0.81,0.81,0.81}
\definecolor{grey82}{rgb}{0.82,0.82,0.82}
\definecolor{grey83}{rgb}{0.83,0.83,0.83}
\definecolor{grey84}{rgb}{0.84,0.84,0.84}
\definecolor{grey85}{rgb}{0.85,0.85,0.85}
\definecolor{grey86}{rgb}{0.86,0.86,0.86}
\definecolor{grey87}{rgb}{0.87,0.87,0.87}
\definecolor{grey88}{rgb}{0.88,0.88,0.88}
\definecolor{grey89}{rgb}{0.89,0.89,0.89}
\definecolor{grey8}{rgb}{0.08,0.08,0.08}
\definecolor{grey90}{rgb}{0.90,0.90,0.90}
\definecolor{grey91}{rgb}{0.91,0.91,0.91}
\definecolor{grey92}{rgb}{0.92,0.92,0.92}
\definecolor{grey93}{rgb}{0.93,0.93,0.93}
\definecolor{grey94}{rgb}{0.94,0.94,0.94}
\definecolor{grey95}{rgb}{0.95,0.95,0.95}
\definecolor{grey96}{rgb}{0.96,0.96,0.96}
\definecolor{grey97}{rgb}{0.97,0.97,0.97}
\definecolor{grey98}{rgb}{0.98,0.98,0.98}
\definecolor{grey99}{rgb}{0.99,0.99,0.99}
\definecolor{grey9}{rgb}{0.09,0.09,0.09}
\definecolor{grey}{rgb}{0.75,0.75,0.75}
\definecolor{honeydew1}{rgb}{0.94,1.00,0.94}
\definecolor{honeydew2}{rgb}{0.88,0.93,0.88}
\definecolor{honeydew3}{rgb}{0.76,0.80,0.76}
\definecolor{honeydew4}{rgb}{0.51,0.55,0.51}
\definecolor{honeydew}{rgb}{0.94,1.00,0.94}
\definecolor{hotpink}{rgb}{1.00,0.41,0.71}
\definecolor{indianred}{rgb}{0.80,0.36,0.36}
\definecolor{ivory1}{rgb}{1.00,1.00,0.94}
\definecolor{ivory2}{rgb}{0.93,0.93,0.88}
\definecolor{ivory3}{rgb}{0.80,0.80,0.76}
\definecolor{ivory4}{rgb}{0.55,0.55,0.51}
\definecolor{ivory}{rgb}{1.00,1.00,0.94}
\definecolor{khaki1}{rgb}{1.00,0.96,0.56}
\definecolor{khaki2}{rgb}{0.93,0.90,0.52}
\definecolor{khaki3}{rgb}{0.80,0.78,0.45}
\definecolor{khaki4}{rgb}{0.55,0.53,0.31}
\definecolor{khaki}{rgb}{0.94,0.90,0.55}
\definecolor{lavenderblush}{rgb}{1.00,0.94,0.96}
\definecolor{lavender}{rgb}{0.90,0.90,0.98}
\definecolor{lawngreen}{rgb}{0.49,0.99,0.00}
\definecolor{lemonchiffon}{rgb}{1.00,0.98,0.80}
\definecolor{lightblue}{rgb}{0.68,0.85,0.90}
\definecolor{lightcoral}{rgb}{0.94,0.50,0.50}
\definecolor{lightcyan}{rgb}{0.88,1.00,1.00}
\definecolor{lightgoldenrod}{rgb}{0.93,0.87,0.51}
\definecolor{lightgoldenrod}{rgb}{0.98,0.98,0.82}
\definecolor{lightgray}{rgb}{0.83,0.83,0.83}
\definecolor{lightgreen}{rgb}{0.56,0.93,0.56}
\definecolor{lightgrey}{rgb}{0.83,0.83,0.83}
\definecolor{lightpink}{rgb}{1.00,0.71,0.76}
\definecolor{lightsalmon}{rgb}{1.00,0.63,0.48}
\definecolor{lightsea}{rgb}{0.13,0.70,0.67}
\definecolor{lightsky}{rgb}{0.53,0.81,0.98}
\definecolor{lightslate}{rgb}{0.47,0.53,0.60}
\definecolor{lightslate}{rgb}{0.47,0.53,0.60}
\definecolor{lightslate}{rgb}{0.52,0.44,1.00}
\definecolor{lightsteel}{rgb}{0.69,0.77,0.87}
\definecolor{lightyellow}{rgb}{1.00,1.00,0.88}
\definecolor{limegreen}{rgb}{0.20,0.80,0.20}
\definecolor{linen}{rgb}{0.98,0.94,0.90}
\definecolor{magenta1}{rgb}{1.00,0.00,1.00}
\definecolor{magenta2}{rgb}{0.93,0.00,0.93}
\definecolor{magenta3}{rgb}{0.80,0.00,0.80}
\definecolor{magenta4}{rgb}{0.55,0.00,0.55}
\definecolor{magenta}{rgb}{1.00,0.00,1.00}
\definecolor{maroon1}{rgb}{1.00,0.20,0.70}
\definecolor{maroon2}{rgb}{0.93,0.19,0.65}
\definecolor{maroon3}{rgb}{0.80,0.16,0.56}
\definecolor{maroon4}{rgb}{0.55,0.11,0.38}
\definecolor{maroon}{rgb}{0.69,0.19,0.38}
\definecolor{mediumaquamarine}{rgb}{0.40,0.80,0.67}
\definecolor{mediumblue}{rgb}{0.00,0.00,0.80}
\definecolor{mediumorchid}{rgb}{0.73,0.33,0.83}
\definecolor{mediumpurple}{rgb}{0.58,0.44,0.86}
\definecolor{mediumsea}{rgb}{0.24,0.70,0.44}
\definecolor{mediumslate}{rgb}{0.48,0.41,0.93}
\definecolor{mediumspring}{rgb}{0.00,0.98,0.60}
\definecolor{mediumturquoise}{rgb}{0.28,0.82,0.80}
\definecolor{mediumviolet}{rgb}{0.78,0.08,0.52}
\definecolor{midnightblue}{rgb}{0.10,0.10,0.44}
\definecolor{mintcream}{rgb}{0.96,1.00,0.98}
\definecolor{mistyrose}{rgb}{1.00,0.89,0.88}
\definecolor{moccasin}{rgb}{1.00,0.89,0.71}
\definecolor{navajowhite}{rgb}{1.00,0.87,0.68}
\definecolor{navyblue}{rgb}{0.00,0.00,0.50}
\definecolor{navy}{rgb}{0.00,0.00,0.50}
\definecolor{oldlace}{rgb}{0.99,0.96,0.90}
\definecolor{olivedrab}{rgb}{0.42,0.56,0.14}
\definecolor{orange1}{rgb}{1.00,0.65,0.00}
\definecolor{orange2}{rgb}{0.93,0.60,0.00}
\definecolor{orange3}{rgb}{0.80,0.52,0.00}
\definecolor{orange4}{rgb}{0.55,0.35,0.00}
\definecolor{orangered}{rgb}{1.00,0.27,0.00}
\definecolor{orange}{rgb}{1.00,0.65,0.00}
\definecolor{orchid1}{rgb}{1.00,0.51,0.98}
\definecolor{orchid2}{rgb}{0.93,0.48,0.91}
\definecolor{orchid3}{rgb}{0.80,0.41,0.79}
\definecolor{orchid4}{rgb}{0.55,0.28,0.54}
\definecolor{orchid}{rgb}{0.85,0.44,0.84}
\definecolor{palegoldenrod}{rgb}{0.93,0.91,0.67}
\definecolor{palegreen}{rgb}{0.60,0.98,0.60}
\definecolor{paleturquoise}{rgb}{0.69,0.93,0.93}
\definecolor{paleviolet}{rgb}{0.86,0.44,0.58}
\definecolor{papayawhip}{rgb}{1.00,0.94,0.84}
\definecolor{peachpuff}{rgb}{1.00,0.85,0.73}
\definecolor{peru}{rgb}{0.80,0.52,0.25}
\definecolor{pink1}{rgb}{1.00,0.71,0.77}
\definecolor{pink2}{rgb}{0.93,0.66,0.72}
\definecolor{pink3}{rgb}{0.80,0.57,0.62}
\definecolor{pink4}{rgb}{0.55,0.39,0.42}
\definecolor{pink}{rgb}{1.00,0.75,0.80}
\definecolor{plum1}{rgb}{1.00,0.73,1.00}
\definecolor{plum2}{rgb}{0.93,0.68,0.93}
\definecolor{plum3}{rgb}{0.80,0.59,0.80}
\definecolor{plum4}{rgb}{0.55,0.40,0.55}
\definecolor{plum}{rgb}{0.87,0.63,0.87}
\definecolor{powderblue}{rgb}{0.69,0.88,0.90}
\definecolor{purple1}{rgb}{0.61,0.19,1.00}
\definecolor{purple2}{rgb}{0.57,0.17,0.93}
\definecolor{purple3}{rgb}{0.49,0.15,0.80}
\definecolor{purple4}{rgb}{0.33,0.10,0.55}
\definecolor{purple}{rgb}{0.63,0.13,0.94}
\definecolor{red1}{rgb}{1.00,0.00,0.00}
\definecolor{red2}{rgb}{0.93,0.00,0.00}
\definecolor{red3}{rgb}{0.80,0.00,0.00}
\definecolor{red4}{rgb}{0.55,0.00,0.00}
\definecolor{red}{rgb}{1.00,0.00,0.00}
\definecolor{rosybrown}{rgb}{0.74,0.56,0.56}
\definecolor{royalblue}{rgb}{0.25,0.41,0.88}
\definecolor{saddlebrown}{rgb}{0.55,0.27,0.07}
\definecolor{salmon1}{rgb}{1.00,0.55,0.41}
\definecolor{salmon2}{rgb}{0.93,0.51,0.38}
\definecolor{salmon3}{rgb}{0.80,0.44,0.33}
\definecolor{salmon4}{rgb}{0.55,0.30,0.22}
\definecolor{salmon}{rgb}{0.98,0.50,0.45}
\definecolor{sandybrown}{rgb}{0.96,0.64,0.38}
\definecolor{seagreen}{rgb}{0.18,0.55,0.34}
\definecolor{seashell1}{rgb}{1.00,0.96,0.93}
\definecolor{seashell2}{rgb}{0.93,0.90,0.87}
\definecolor{seashell3}{rgb}{0.80,0.77,0.75}
\definecolor{seashell4}{rgb}{0.55,0.53,0.51}
\definecolor{seashell}{rgb}{1.00,0.96,0.93}
\definecolor{sienna1}{rgb}{1.00,0.51,0.28}
\definecolor{sienna2}{rgb}{0.93,0.47,0.26}
\definecolor{sienna3}{rgb}{0.80,0.41,0.22}
\definecolor{sienna4}{rgb}{0.55,0.28,0.15}
\definecolor{sienna}{rgb}{0.63,0.32,0.18}
\definecolor{skyblue}{rgb}{0.53,0.81,0.92}
\definecolor{slateblue}{rgb}{0.42,0.35,0.80}
\definecolor{slategray}{rgb}{0.44,0.50,0.56}
\definecolor{slategrey}{rgb}{0.44,0.50,0.56}
\definecolor{snow1}{rgb}{1.00,0.98,0.98}
\definecolor{snow2}{rgb}{0.93,0.91,0.91}
\definecolor{snow3}{rgb}{0.80,0.79,0.79}
\definecolor{snow4}{rgb}{0.55,0.54,0.54}
\definecolor{snow}{rgb}{1.00,0.98,0.98}
\definecolor{springgreen}{rgb}{0.00,1.00,0.50}
\definecolor{steelblue}{rgb}{0.27,0.51,0.71}
\definecolor{tan1}{rgb}{1.00,0.65,0.31}
\definecolor{tan2}{rgb}{0.93,0.60,0.29}
\definecolor{tan3}{rgb}{0.80,0.52,0.25}
\definecolor{tan4}{rgb}{0.55,0.35,0.17}
\definecolor{tan}{rgb}{0.82,0.71,0.55}
\definecolor{thistle1}{rgb}{1.00,0.88,1.00}
\definecolor{thistle2}{rgb}{0.93,0.82,0.93}
\definecolor{thistle3}{rgb}{0.80,0.71,0.80}
\definecolor{thistle4}{rgb}{0.55,0.48,0.55}
\definecolor{thistle}{rgb}{0.85,0.75,0.85}
\definecolor{tomato1}{rgb}{1.00,0.39,0.28}
\definecolor{tomato2}{rgb}{0.93,0.36,0.26}
\definecolor{tomato3}{rgb}{0.80,0.31,0.22}
\definecolor{tomato4}{rgb}{0.55,0.21,0.15}
\definecolor{tomato}{rgb}{1.00,0.39,0.28}
\definecolor{turquoise1}{rgb}{0.00,0.96,1.00}
\definecolor{turquoise2}{rgb}{0.00,0.90,0.93}
\definecolor{turquoise3}{rgb}{0.00,0.77,0.80}
\definecolor{turquoise4}{rgb}{0.00,0.53,0.55}
\definecolor{turquoise}{rgb}{0.25,0.88,0.82}
\definecolor{violetred}{rgb}{0.82,0.13,0.56}
\definecolor{violet}{rgb}{0.93,0.51,0.93}
\definecolor{wheat1}{rgb}{1.00,0.91,0.73}
\definecolor{wheat2}{rgb}{0.93,0.85,0.68}
\definecolor{wheat3}{rgb}{0.80,0.73,0.59}
\definecolor{wheat4}{rgb}{0.55,0.49,0.40}
\definecolor{wheat}{rgb}{0.96,0.87,0.70}
\definecolor{whitesmoke}{rgb}{0.96,0.96,0.96}
\definecolor{white}{rgb}{1.00,1.00,1.00}
\definecolor{yellow1}{rgb}{1.00,1.00,0.00}
\definecolor{yellow2}{rgb}{0.93,0.93,0.00}
\definecolor{yellow3}{rgb}{0.80,0.80,0.00}
\definecolor{yellow4}{rgb}{0.55,0.55,0.00}
\definecolor{yellowgreen}{rgb}{0.60,0.80,0.20}
\definecolor{yellow}{rgb}{1.00,1.00,0.00}

\usepackage{ulem}       

\begin{document}        


\maketitle
\label{firstpage}

\begin{abstract}
We present the largest, most homogeneous catalogue of merging galaxies in the nearby universe obtained through the Galaxy Zoo project - an interface on the world-wide web enabling large-scale morphological classification of galaxies through visual inspection of images from the Sloan Digital Sky Survey (SDSS). The method converts a set of visually-inspected classifications for each galaxy into a single parameter (the `weighted-merger-vote fraction,' $f_m$) which describes our confidence that the system is part of an ongoing merger. We describe how $f_m$ is used to create a catalogue of 3003 visually-selected pairs of merging galaxies from the SDSS in the redshift range $0.005 < z <0.1$. We use our merger sample and values of $f_m$ applied to the SDSS Main Galaxy Spectral sample (MGS) to estimate that the fraction of volume-limited ($M_r < -20.55$) major mergers ($1/3 < \mbox{M}^*_1/\mbox{M}^*_2 < 3$) in the nearby universe is $1 - 3 \times C\%$ where $C \sim 1.5$ is a correction factor for spectroscopic incompleteness. Having visually classified the morphologies of the constituent galaxies in our mergers, we find that the spiral-to-elliptical ratio of galaxies in mergers is higher by a factor $\sim 2$ relative to the global population. In a companion paper, we examine the internal properties of these merging galaxies and conclude that this high spiral-to-elliptical ratio in mergers is due to a longer time-scale over which mergers with spirals are detectable compared to mergers with ellipticals.
\end{abstract}


\section{Introduction}

\label{intro}

Galaxy mergers and interactions are connected to many pressing questions concerning the origin, evolution and properties of cosmic structures. These issues include the formation of galaxies (\citealt{white}; \citealt{lacey}; \citealt{con2006}; \citealt{delucia}), environmental effects on morphology (\citealt{capak}; \citealt{park}; \citealt{ball}; \citealt{vdw}) and the distribution of the dark matter that drives the merger process (\citealt{bond}; \citealt{cole}; \citealp{fak}). At the sub-galactic scale mergers have been invoked to explain a variety of observations, notably localised bursts of star formation (\citealt{kennicutt1}; \citealt{schweizer2}; \citealt{dimatteo}; \citealt{woods}; \citealt{barton}; \citealt{li}; \citealt{cox}) and induced nuclear activity (\citealt{keel}; \citealt{schawinski2}; \citealt{jog2}). The far-reaching effects that mergers are thought to produce makes their empirical examination an important task.   

To date though, such studies have concentrated mostly on merger {\it rates} (\citealt{carlberg}; \citealt{lefevre}; \citealt{patton2002}; \citealt{con2003b}; \citealt{bundy}; \citealt{bell}; \citealt{con2008}; \citealt{lin}; \citealt{lotz2}; \citealt{hsieh}, \citealt{patton2008}),  with comparatively little carried out to examine their morphologies and internal properties (though see \citealt{li}; \citealt{ellison}) for reasons to be discussed (\S \ref{pastmethods}). This is unfortunate - to understand the exact role played by mergers in galaxy evolution, it is inadequate to measure their rates alone since the processes that determine the morphological outcomes of mergers are still not fully understood. 

It has been widely believed since the simulations of \citet{toomre} that two spirals {\it can} merge to form an elliptical but this does not mean that they {\it must}. On the contrary, recent studies have argued convincingly that, in some cases at least, disc galaxies are able to survive (multiple) major mergers \citep{hopkins}. Where this does happen, the probability of disc survival must be assumed {\it a priori} to depend on the properties of its progenitors such as environment, gas content and feedback mechanisms. These in turn correlate with galactic morphology. There is also good reason to think that the time-scales over which mergers are detectable depends on the internal properties of the progenitors (\citealp{lotz3}). Ideally then, calculations that convert merger fractions into merger rates as part of some hierarchical-structure scheme (such as modeled by \citealp{khochfar2}) should take the properties and morphologies of their sample into account. Only then can models of hierarchical galaxy formation be fully tested.

The Galaxy Zoo project contributes to this important task by providing a snapshot of mergers as they appear in the local universe (in this study we focus on the range $0.005<z<0.1$). Plans are also underway to apply the same merger-locating system to higher redshift surveys. In this paper we present a simple but powerful method for identifying mergers using the world-wide web and a robust sample of 3003 merging systems obtained by it (\S \ref{catalogue}).\footnote{We have labeled this catalogue GZM1 (Galaxy-Zoo Mergers 1).} We then discuss three important ratios revealed by this study: the merger fraction of the local universe (\S \ref{aaa} and \ref{aaa2}), the fraction of spectral pairs in merging systems in SDSS (relevant to discussions for the close-pairs technique; \S \ref{close_pairs}) and the spiral-to-elliptical ratio of galaxies that are in merging systems (\S \ref{s2e}). In the companion paper, \citealt{darg} (D09b from here on), we present the environment and internal properties of these merging galaxies. 

Galaxy Zoo is a new player in the game of merger-location techniques. Before describing our results, we begin with a description of current methods used to find mergers so that the role and value of Galaxy Zoo can be understood in the context of contemporary research.

\begin{figure}
	\includegraphics[width=84mm]{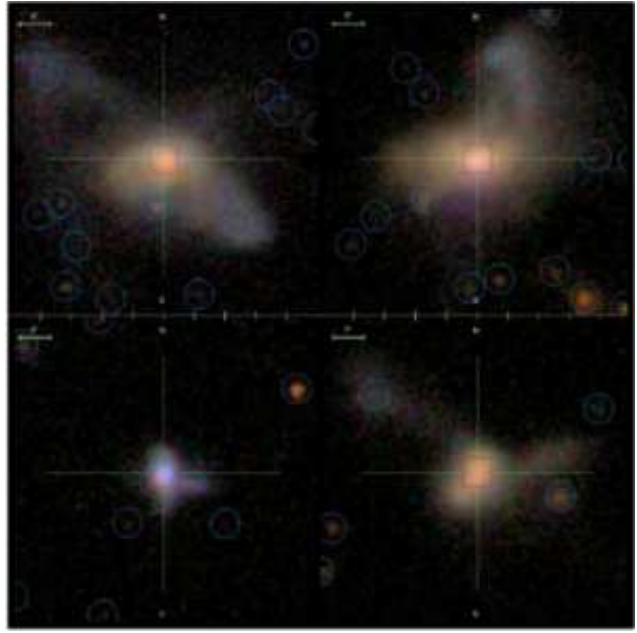}
	\caption[Figure 4]{Example images of `strongly-perturbed' systems which are not deblended into two photometric objects that plausibly represent the photometry of the progenitors. These are usually late-stage mergers and cannot be detected by the close-pairs technique. {\color{black}Blue circles mark the centre of SDSS photometric objects. Red squares mark the centre of SDSS spectral objects. The scale bar is $5''$ for each example image.} }
	\label{iexample}
\end{figure}

\subsection{Locating Mergers: Past Methods }
\label{pastmethods}

Visual inspection remains the most trustworthy way of determining whether or not a galaxy is merging. However, the advent of large galaxy surveys such as SDSS (\citealt{york}) involving $\sim10^{6}$ objects has rendered classification by individual researchers impractical. Automated methods have therefore been developed to approximate the human decision making process but, being only approximations, encounter certain difficulties that Galaxy Zoo can help overcome.

\subsubsection{Close-Pair Statistics}
\label{cps}

Accurate measures of position and redshift in large scale surveys have made straightforward the task of finding close pairs in the (local) universe. However, the peculiar velocities of the individual galaxies can significantly offset their redshift-inferred line-of-sight separation producing spectral pairs of non-interacting systems which may or may not merge. One can apply statistical arguments as to what fraction of close-pairs within some ensemble will soon merge, but no single system can be safely assumed to be merging without cross-examination. The method is best used, therefore, to estimate merger {\it rates} parametrised by the {\color{black}convention: {\it merger rate}  $\sim(1+z)^m$.} 

Another limitation of close-pairs methods in surveys such as SDSS is the requirement that {\it both} galaxies have spectra. This requires a delicate act of deblending by the data-reduction pipeline. SDSS will convert the photometry of an extended body like a galaxy (the `parent') into individual photometric `objects' (the `children') depending on how many peaks are present in the blended image (\citealt{stoughton}). Iterations are performed allotting different magnitudes to the child-objects until an optimal distribution is attained so that all children sum to give the total flux of the parent. 

Any photometric-child object with Petrosian magnitude $r<17.77$ is designated as a spectral target by the SDSS pipeline. It follows that only when a contiguous body, such as two interacting galaxies, is deblended into (at least) two photometric objects both having $r<17.77$ can it contain two spectral targets. Therefore mergers with only one deblended object of $r<17.77$ cannot be identified by the close-pairs technique (\citealt{strauss}; \citealt{patton2008}). This will occur for many minor mergers or for late stage mergers where the cores have drawn close to each other and the pipeline reads them as a single peak. Figure \ref{iexample} shows examples of strongly-perturbed systems with only one spectral target. 

Even when two objects are deblended in a contiguous image and registered as spectral targets - relatively few systems will obtain spectra on {\it both} objects due to fiber collisions within the spectrometer. Two SDSS spectroscopic targets cannot acquire spectra simultaneously if they are within $55''$ of each other and so only systems contained within `tile overlap regions' can have spectral objects with an angular separation less than this (\citealt{strauss}, \citealt{blanton2}). Only about $\sim30\%$ of the SDSS sky rests within overlap regions thus limiting the sample size of such systems. 

\begin{figure}
	\includegraphics[width=84mm]{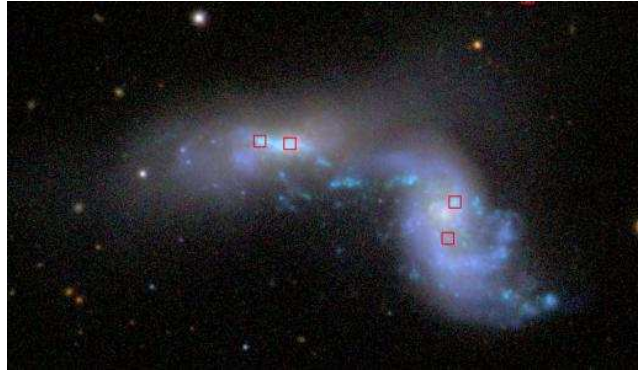}
	\caption[Figure 1b]{Example image of binary merger with four spectral objects {\color{black}(red squares)}. It is difficult for `blind' close-pairs techniques to distinguish this from a pair of binary mergers (or a four-way merger). }
	\label{4spec_merger}
\end{figure}

Conversely, a system deblended into multiple targets, as in Figure \ref{4spec_merger}, can acquire too many spectral objects so that a single galaxy might appear as a close-pair or a binary merger might appear as a multi-merger. Close-pairs studies typically limit the redshift and magnitude ranges in order to avoid this effect. \citet{woods}, for example, limit their close-pairs sample to $\sim0.027<z<0.17$ and require $>20\%$ of the galaxy's total flux land within the fiber aperture in order to exclude very extended nearby galaxies. They still find though that $\sim15\%$ of their minor-pairs catalogue are false pairs after visual inspection.

To summarise, without visual cross-examination, close-pairs methods are prone to include false pairs and non-gravitationally bound systems and are best used, therefore, to estimate merger {\it rates} via application to large ensembles after estimations for contamination and incompleteness are taken into account. To examine the {\it properties} of merger systems, one should robustly identify {\it actual} mergers as opposed to claims that some fraction of an ensemble is `likely' to be merging. 

\subsubsection{Automated Quantification of Morphological Disturbance}

Pattern recognition techniques applied to galaxy images pose a formidable programming challenge. Background noise and contaminants first need to be removed, then an automated technique needs to quantify how `disturbed' a morphology is independent of viewing angle. This might not be so difficult if it were not the case that unperturbed galaxies themselves vary so much from the structurally simple (ellipticals) to the complex (spirals). In particular, filtering asymmetric systems like mergers from spirals with extended star-formation requires great finesse.
  
Nonetheless, promising automated techniques have been developed in recent years that generate parameters related to morphology. A common technique has been to measure concentration (C), asymmetry (A) and, in some cases, clumpiness/smoothness (S). One then partitions  CA(S) space into morphological categories and identifies an object as a merger if it lies within the designated sub-volume (see \citealt{con2003a} and \citealt{con2003b} on the use of the `CAS' system to locate mergers). 

Another pair of parameters used in this fashion are the Gini-coefficient (G) (\citealt{abraham}) and second-order moment of the brightest $20\%$ of the galaxy's light ($\mathrm{M}_{20}$), most notably by \citet{lotz2}. (For a concise description of how to calculate C, A, S, G and $\mathrm{M}_{20}$ see \citealt{con2008}.) It is argued in \citet{lotz1} that this technique operates better at low signal-to-noise ratios and is more sensitive to late-stage morphologies than the CAS system (though see \citealt{lisker} for a critique of its effectiveness). More recent studies have combined and compared the two techniques (e.g. \citealt{scarlata}; \citealt{con2008}; \citealt{lotz3}). Artifical Neural Nets are another promising technique for identifying morpholgies (\citealt{lahav}, \citealt{ball2}) though they have not yet been applied to merger studies specifically.

CAS and G$\mathrm{M}_{20}$ remain effective at high redshifts so long as the image quality remains high and, since they are automated, can process images quickly once the pipeline has been established. However, these techniques also require visual examination to cross-check results and to fine-tune the partition boundaries that define a merger. 

Systematic uncertainties usually remain. \citet{jog1} recently report that the CAS criterion failed to pick up $37-58\%$ of their visually classified ``strongly disturbed'' morphologies while including a ``significant number of relatively normal galaxies'' - an effect that was predicted by simulations in \citet{con2006b}. \citet{con2008} recently classified 993 images from the Hubble-Ultra-Deep Field by visual inspection, the CAS volume and the G$\mathrm{M}_{20}$ area. They found some notable disagreements, in particular, some systems identified as mergers by their location in G$\mathrm{M}_{20}$ space are not identified as mergers using the CAS system and vice-versa (see for example Figures 8 and 9 plus captions). Within the $0.4<z<0.8$ range only $44\pm6\%$ of the galaxies mapped into the G$\mathrm{M}_{20}$ merger region were visually classified as `peculiar.' 

Recently \citet{lotz3} performed an extensive study on the merger-detection sensitivity of the CA, G$\mathrm{M}_{20}$ and close-pairs techniques using simulations. They find that C, A, G and $\mathrm{M}_{20}$ methods are only sensitive to mergers at specific stages of the process, particularly the first pass and final coalescence of the galaxies. The study also confirms that the merger time-scales and parameters (such as gas-fractions, pericentric distance and relative orientation), for which these three techniques remain sensitive, differ significantly. 

Evidently then, much work remains before we can confidently do away with visual inspection of images in merger studies.\footnote{``Men trust their ears less than their eyes.'' Herodutus, \\{\it The Histories}, 5th c. B.C., 1.8, 
trans D. Godley.} 

\subsubsection{Visual Inspection by Research Groups}

In the pre-digital age, surveys of morphological classification by visual inspection were mostly limited to studies based on cluster images (\citealt{oemler}, \citealt{dressler}). With the modern development of large surveys and high resolution imaging, the potential to extract large samples of morphologically classified galaxies in differing environments has vastly improved. The largest visual classification project by a research group is \citet{schawinski2} who visually examined 48,023 SDSS galaxies in compiling the MOSES catalogue (MOrphologically Selected Ellipticals in SDSS). They found a significant blue-population that had been excluded by studies such as \citet{bernardi} which identified ellipticals by their presumed properties. Selecting morphologies by {\it a priori} assumptions is often pragmatic and necessary but ultimately begs the question as to what the properties of a given morphology actually are. Unless we can be certain that a set of properties maps one-to-one with morphology we must continue with visual examination.

Studies to select {\it mergers} by visual-selection have not reached such scales as MOSES until this work. \citet{lefevre} visually examined 285 Hubble images and found a merger fraction of $10\pm2\%$ over $0<z\la1.2$. \citet{nakamura} visually classified 2418 images of SDSS galaxy objects with Petrosian $r<16.0$ (i.e. low redshift) finding that 35/1875 of their morphological classifications were Im (`highly irregular' following Hubble's notation). A similar study by the same group, \citet{fuk}, visually classified 2658 images from SDSS with Petrosian $r<15.9$ and found that $1.5\%-1.7\%$ of these nearby magnitude limited galaxies were `interacting.' 

Thus, until now, merger studies have been limited to catalogues of $\sim10^{3}$ galaxies due to the time-consuming and monotonous nature of the task. By contrast Galaxy Zoo allows us to acquire effective visual classification for morphologies and mergers for samples of $\sim10^{5}$ galaxies with relative ease.

\subsection{Locating Mergers with Galaxy Zoo}

Galaxy Zoo (GZ) is a user interface on the world-wide web\footnote{http://www.galaxyzoo.org/} drawing upon images from SDSS. Volunteers from the public are instructed and commisioned with the collective task of visual classification of $\sim900,000$ galaxies from SDSS DR6. A complete description of the project including design details and initial data reduction is given by \citet{lintott1}. It is also shown that public users are, in large numbers, about as good as `experts' at identifying morphologies.

The project has proved to be a tremendous success. To date, over 140,000 volunteers have participated and collectively offered classifications for all $\sim900,000$ images, on average, fifty times over. Data from the project has already been employed to find the statistical properties of spiral-galaxy-spin orientation (\citealt{land}; \citealt{slosar}), the relationships between environment, morphology and colour (\citealt{bamford}; \citealt{skibba}), to study optically blue early-type galaxies at very low redshift (\citealt{schawinski4}) and has also lead to some serendipitous discoveries (\citealt{lintott3}; \citealt{cardamone}). This work concentrates on the results of the merger-location functionality of Galaxy Zoo whose details we now describe. 

\begin{figure*}
	\includegraphics[width=140mm]{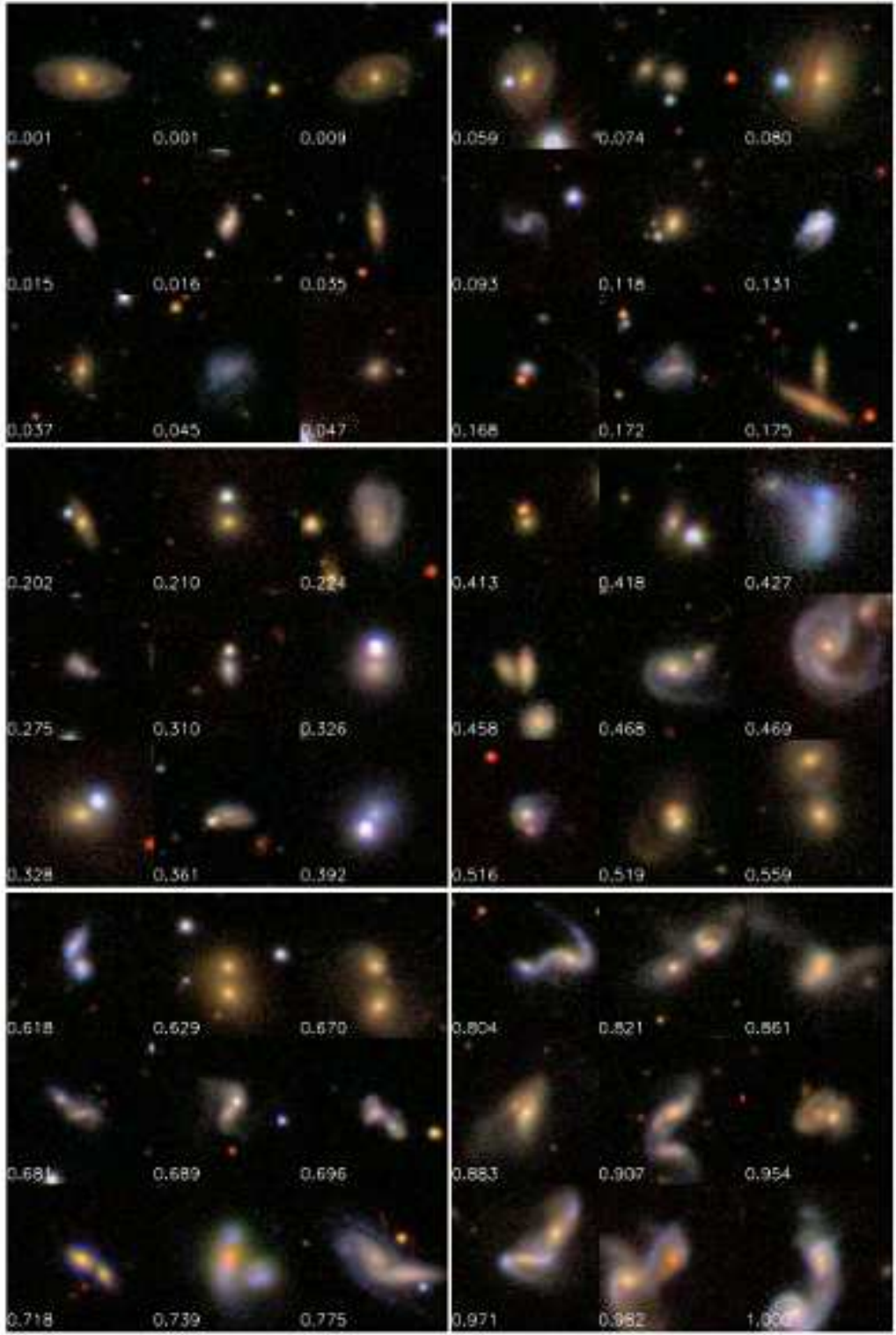}
	\caption[Figure 1]{Example images of prospective merger systems. The number given for each image is its weighted-merger-vote fraction, $f_m$. The panels of nine images are grouped into the following bins: top-left: $0<f_m\leq0.05$, top-right: $0.05<f_m\leq0.2$; middle-left: $0.2<f_m\leq0.4$, middle-right: $0.4<f_m\leq0.6$; bottom-left: $0.6<f_m\leq0.8$, bottom-right: $0.8<f_m\leq1.0$. Our catalogue is made up of mergers with $f_m>0.4$. Each image tile is $40''\times 40''$.}	
	\label{collation}
\end{figure*}


\section{The Galaxy Zoo Data}
\label{sec2}
\subsection{Constructing a Merging-Pairs Catalogue}
\label{catalogue}
Galaxy-Zoo users are asked to classify an SDSS target as 

\begin{enumerate}
 	\item elliptical (e)
 	\item spiral (s)\footnote{More specifically, users are asked to specify whether a spiral galaxy is rotating clockwise, anti-clockwise or is edge-on/unclear.}
	\item star/bad image (b), or
	\item merger (m).
\end{enumerate}

The Galaxy Zoo project is the first of its kind and so, not knowing how well it would be taken up by the public, the design of the interface prioritised simplicity over detail. A single button labeled `merger' is all that was offered. 

\begin{figure}
	\includegraphics[width=84mm]{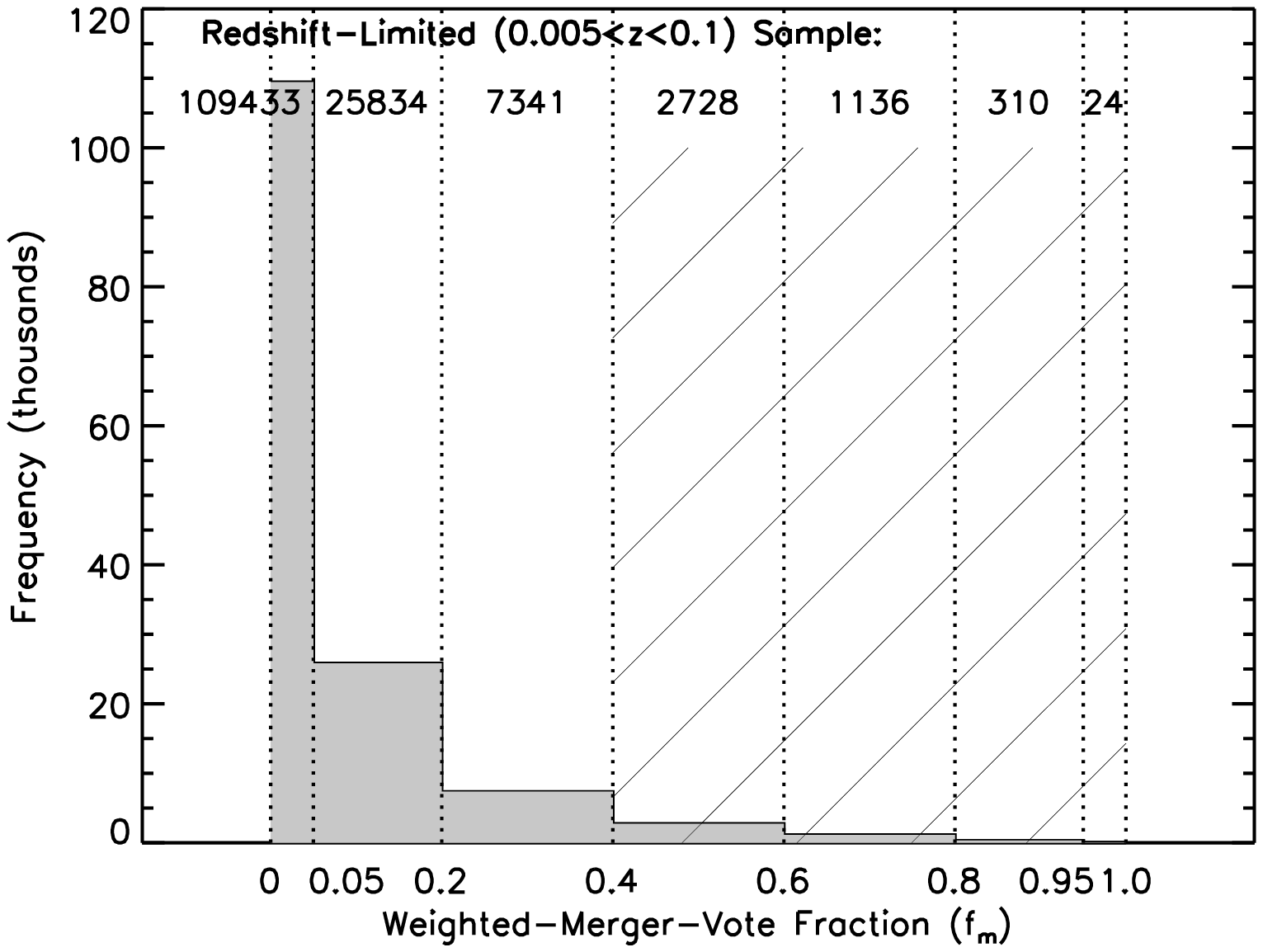}
	\caption[Figure 2]{The distribution of weighted-merger-vote fractions in the Galaxy Zoo database for objects with spectra for $0.005<z<0.1$. From these we use objects with $0.4<f_m\leq1.0$ to construct our merging-pairs catalogue (cross-hatched). There are $157,376$ objects with $f_m=0$ exactly. The numbers on the graph show the occupancy of each bin (e.g. there are $109,433$ with $0<f_m \leq 0.05$). In total, 304,182 Galaxy Zoo spectral objects lie between $0.005<z<0.1$.}
	\label{votes_40_100}
\end{figure}

The raw data we use to build our catalogue is what we call the {\it weighted-merger-vote fraction} ($f_m$).\footnote{Similar parameters are calculated for the other categories such that $f_e + f_s + f_b + f_m = 1$.} GZ has obtained a value for this parameter for $893,292$ SDSS galaxies from DR6.\footnote{This is the same parent population obtained after six-months of running Galaxy Zoo that is used in \citet{lintott1}, \citet{land}, \citet{bamford} and \citet{schawinski4}.} The $f_m$ values are calculated by taking the number of {\it merger} classifications ($n_{m}$) for a given GZ object and dividing it by the {\it total} number of classifications ($n_{e,s,b,m}$) for that object multiplied by a weighting factor $W$ that measures the quality of the particular users that have assessed the object. The quality of an individual user is determined by measuring to what extent that person agrees with the majority opinion for all objects the individual has viewed. The weighting factor, $W$, thereby represents all the iterations carried out by equations (1) \& (2) of \citealt{lintott1}).\footnote{$W$ here is {\it not} the $w_k$ of \citealt{lintott1}. $w_k$ is the weighting of each individual user whereas $W$ represents their combined weighting for each individual GZ object.}  

\begin{center}\begin{equation}\label{f_m}f_{m}=\frac{Wn_{m}}{n_{e,s,b,m}}.\end{equation}\end{center} 

The parameter $f_m$ ranges from 0 to 1 so that an object with $f_m=0$ should look nothing like a merger and $f_m=1$ should look unmistakably so. Figure \ref{collation} shows some example images taken from GZ labeled by their $f_m$ values. It is interesting to see how low $f_m$ is when these images start to look, at least superficially, like mergers ($f_m\sim0.2-0.4$). Evidently, users were rather conservative in calling something a merger.

Figure \ref{votes_40_100} shows the distribution of $f_m$ for the entire GZ catalogue. In building our first merger catalogue we determined a cut-off point for $f_m$ above which most systems will be proper galactic mergers and involve a sample size manageable by a research team. By comparing the merger-vote fraction with images like those in Figure \ref{collation}, we decided to build our first catalogue using only systems with $f_m>0.4$ {\color{black} and only using GZ objects with spectra whose spectroscopic redshift lies in the range $0.005<z<0.1$. SDSS spectroscopic targets are selected for galaxies with apparent magnitude $r<17.77$ which corresponds to $M_r<-20.55$ at $z=0.1$.} This absolute magnitude corresponds to a minimum stellar mass of $\sim10^{10}\mbox{M}_{\astrosun}$ (see Figure \ref{scatter_kmass}) so that our upper limit ($z=0.1$) of our volume-limited sample will be inclusive of intermediate size galaxies.\footnote{{\color{black} For clarity, the merger catalogue that we construct is {\it not} volume-limited, i.e. the systems are red-shift limited ($0.005<z<0.1$) but the absolute magnitude ($M_r$) is subject to no formal constraint in order for that object to belong to the catalogue, although its apparent magnitude must have been such that the SDSS pipeline deblended the system into a spectral target (with $r<17.77$). For certain investigations, however, it is important to impose an absolute-magnitude cut of $M_r<-20.55$ on the catalogue in order to ensure completeness across the redshift range we are using. We state when we do this and refer to such a subset as a `volume-limited' sample.}} The resolution of images beyond $z>0.1$ is rapidly diminishing which would lead to unreliable visual classifications of morphology. 

The lower limit $z=0.005$ is to minimise the number of mergers that go undetected due to an incomplete field of view.\footnote{This arises because GZ images are scaled for viewing according to the Petrosian radius of the object's model magnitude. The photometry of very large and close-by galaxies is often deblended into multiple SDSS objects. The more deblended objects there are, the less the Petrosian radius of any single deblended object represents the galaxy as a whole and this brings about an inappropriately small image-scale for viewing some close-by galaxies. Such systems would be viewed by users as nothing but a galaxy core and not, consequently, voted as a merger.} These cases are rare since the number of SDSS objects peaks near $z=0.08$ and only a few percent have $z<0.02$.  

{\color{black} After applying the cut to only those objects with spectral redshifts between $0.005<z<0.1$, we have $304,182$ objects (see Figure \ref{votes_40_100}).
To find mergers within this set, we apply the cut $0.4<f_m\leq1.0$ leaving 4198 GZ objects with spectra.} We exclude those SDSS targets which are yet to acquire spectra as we desire accurately measured redshifts.\footnote{Mergers are inherently `messy' systems and so photometric redshifts calculated by comparison to standard templates are prone to error. In fact we found the {\it mean} absolute difference between all the photometric and spectroscopic redshifts in our merger sample to be $\sim0.051$ - more than half of our redshift range!} 

Although the purpose of the Galaxy Zoo project is to significantly reduce the need for research teams to visually inspect large catalogues, it is still necessary at this stage to double check the results. Visual re-examination of the sample by our group allowed us to  

\begin{enumerate}
 	\item remove any non-merging systems,
 	
 	\item visually select an appropriate SDSS object to represent the merging partner, and
	\item assign morphologies to the galaxies in each merging system.
\end{enumerate}

We briefly describe our methods for each of these tasks.

	
\label{flow}

\subsection{Removing Non-Merging Systems}

The examples of Figure \ref{collation} demonstrate that the weighted-merger-vote fraction is strongly correlated with how `merger-like' an image appears to be. The outcomes of GZ are therefore similar in effect to automated methods like CAS and G$\mathrm{M}_{20}$ which also map images to parameter spaces. In our case though $f_m$ is a single parameter (making it easier to divide up `merger' and `non-merger' zones) and, by utilising the pattern-recognition capacity of many human minds, overcomes the need to remove background noise, recognise anomalies, etc. We find that for $f_m\ga0.6$, all systems are robust mergers. However, three causes for mis-classification begin to emerge as $f_m$ decreases and become common for $f_m\la0.4$:

\begin{enumerate}

\item projection of {\bf galaxies} along the line of sight, 

\item projection of nearby {\bf stars} onto distant galaxies and 

\item cases which are `border-line' mergers.

\end{enumerate}

Galactic projections occur when two galaxies have similar celestial coordinates but are separated by a significant radial distance. We can easily spot projections when {\it both} galaxies have spectral redshifts. However, many of our candidate systems have only one redshift. Spotting a projection in such cases can only be done through visual examination of the image for signs of interaction. Our choice to use only systems with $f_{m}>0.4$ meant, however, that the need for such difficult decisions was rare. 

Stellar projections were the more common problem. We were able to eliminate stars easily from our sample though by their characteristic point spread function (PSF) using the associated parameter `type' from the SDSS {\texttt PhotoTag} table. This is a discrete label given to every photometric object in the SDSS database that indicates whether an object is `point-like' or `extended' and can, on this basis, reliably determine ($>98\%$) whether a luminous object is a star or not \citep{strauss}. 

`Border-line' cases involve galaxies that {\it are} morphologically disturbed but do not necessarily merit the term `merger.' All galaxies are merging with {\it something} which can range from molecules (accretion) to a small galaxy (minor merger) to a galaxy roughly its own size (major merger). Deciding where in this spectrum of possibilities the term `merger' becomes appropriate is rather subjective. At this stage, therefore, our strategy was simply to decide by visual inspection whether a given GZ galaxy had a `strongly-perturbed' morphology. By `strongly-perturbed' here we mean that a galaxy was in a morphological state that was unlikely to have occurred without some external interaction. Starting with this inclusive criterion allows us obtain completeness for anything deserving the term `merger.' We found in fact that borderline cases were again rare for systems with $f_m>0.4$. However, difficult decisions regarding projections and border-line cases are abundant for systems where $0.05<f_m\la0.4$ (for example the image of Figure \ref{collation} with $f_m=0.175$) and it will therefore require a great deal of care if we wish to expand our catalogue into the $f_m<0.4$ range in the future.

We then examine each `strongly-perturbed' system with the aim of selecting an SDSS object to represent the body responsible for the disruption caused to the galaxy labeled with $f_m>0.4$ (if any is identifiable). The details of this procedure are discussed below. With photometric objects representing the `perturber' we can distinguish major mergers more objectively and their completeness can be assumed since they are a subset of `strongly-perturbed' systems.

\begin{figure}
	\includegraphics[width=84mm]{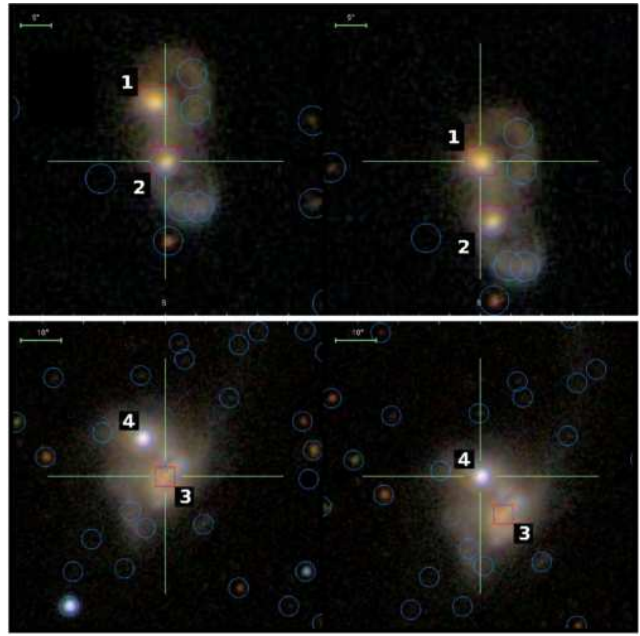}
	\caption[Figure 3]{Images exemplifying the construction of the merging-pairs catalogue. All spectral targets (red boxes) are Galaxy Zoo objects. In the upper panel, we select the two spectral objects (1 and 2) from Galaxy Zoo to represent the merging pair. The merging system in the lower panel only has one spectral target (3) which was found by Galaxy Zoo. We therefore examine all neighbouring SDSS objects in order to select one (in this case 4) to represent the merging partner galaxy. Our final catalogue has 3003 such pairs.}
	\label{selection}
\end{figure}

\subsection{Visual Selection of Merging Partner}
\label{vis_selection}

\begin{figure}
	\includegraphics[width=84mm]{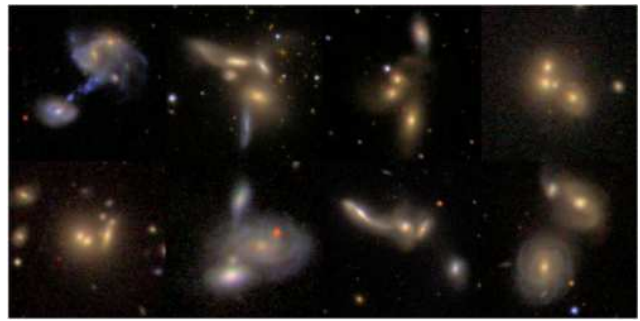}
	\caption[Figure 5]{Example images of multi-merging systems. We generally included these in the binary merger catalogue by selecting the closest and brightest galaxy to accompany the Galaxy-Zoo object. We found 39 systems that could confidently be described as multi-mergers.}
	\label{lcand_collation8}
\end{figure}

As mentioned, in order to create a catalogue of merging pairs, we needed to manually select an appropriate SDSS object as a partner for the object supplied by GZ with $f_m>0.4$. We did this using an IDL routine that allows for the rapid examination of the image and photometry of all objects within $30''$ of the GZ object given by the {\texttt Neighbors} table in the SDSS database. We choose whichever object appears to be the most `plausible' representation of the body responsible for the morphological disruption to the galaxy represented by the GZ object. Plausibility was judged on the basis of object brightness (with brighter objects in the $r$-band being preferable) and visual `common-sense.' See Figure \ref{selection} for examples of this procedure. For most systems, which are binary mergers, this choice was straightforward as they usually have either \\ \\ 
(a) spectra centred on both galaxies or \\
(b) a spectral object centred on one galaxy and only photometry on the other.\\ \\
\indent However, not all `strongly-perturbed' systems appear as simple binary mergers. Additionally, there are cases where \\ \\
(c) galaxies are in the final stages of a merger and its progenitors are no longer distinguished by the SDSS pipeline (`post-mergers'), \\
(d) a galaxy has been perturbed by a close encounter with a neighbour no longer in view (`fly-by') and, occasionally, \\
(e) a merging system involves three or more galaxies.  \\ \\
The wide range of possibilities makes merger taxonomy a difficult task. In constructing a merging-pairs catalogue we therefore proceed as follows for these various cases. 

{\color{black} For case (a) we usually choose these two spectral objects to represent our merging pair (see figure \ref{selection}). If the merger companion does not have spectra then we visually select the best photometric object available to represent the merging partner (case b).  
	
Cases (c) and (d) are sometimes difficult to distinguish and usually occur when only one galaxy core is apparent with the peripheries undergoing extensive tidal disruption. This usually means that no photometric object is available to plausibly represent the perturbing body and so, in such cases, we simply decline to select a merging partner. They remain in the category of `strongly-perturbed' systems (and are included in our calculation of the merger fraction; see \S \ref{aaa} and \S \ref{aaa2}) but are not included in the merging-pairs catalogue. Figure \ref{iexample} shows examples of these two categories. 

In the case of (e), where several galaxies are merging at once, we decided to first note them and then include them in the merging-pairs catalogue by selecting the closest and brightest object to the GZ-supplied galaxy. Figure \ref{lcand_collation8} shows examples of such systems. The catalogue is technically a mix between `binary-mergers' and `multi-mergers' and so we call it the `merging-pairs' catalogue since it contains 3003 pairs of galaxy objects which are all in merging systems.

To summarise, through this refining and pairing process we converted 4198 objects to 681 pairs where {\it both} objects have spectra (case a) and 2322 pairs where only one object has spectra (case b). These make the 3003 pairs of the catalogue. 370 of the 4198 systems were considered strongly-perturbed but unsuitable to be put into a pair (cases c and d). Only 39 of the 3003 pairs are confirmed to be multi-mergers (case e). The remaining 144 objects\footnote{$4198-3003-681-370=144$.} were discarded because they were deemed to be in non-strongly-perturbed systems (mainly stellar overlaps).  }

\subsection{Assigning Morphologies}
\label{assign_morph}
\begin{figure}
	\includegraphics[width=84mm]{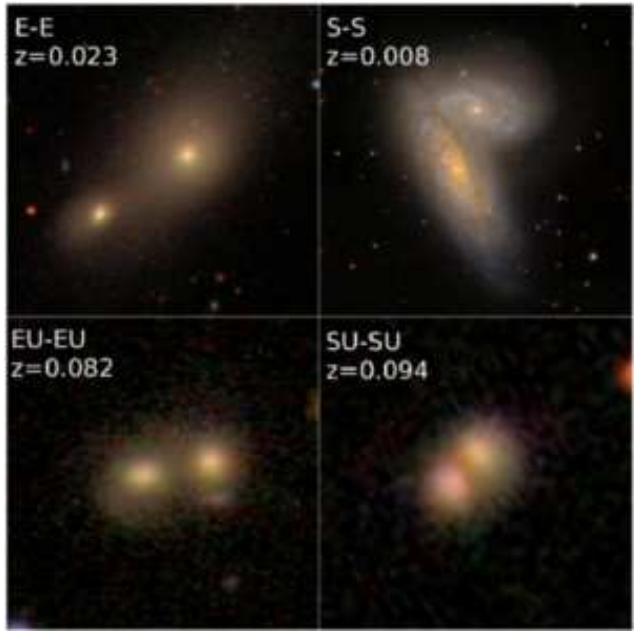}
	\caption[Figure 6]{Example images of galaxies for morphological categories E, S, EU and SU. Any combination of these is possible. Poorer image resolution as $z\rightarrow0.1$ affects certitude of morphological components. The bottom panels show examples of `unsure' morpholgies. If a morphologly is unclear, it will normally be assigned EU type by surface brightness profile if there is a drop off in brightness from the centre. SU types generally have a more uniform surface brightness profile in our approach. {\color{black}The colour of the RBG images was also used as a visual guide} in very difficult cases under the assumption that EU types should be more red. Image widths: E-E $\sim120''$, S-S $\sim240''$, EU-EU $\sim40''$ \& SU-SU $\sim30''$.}
	\label{morphChoice}
\end{figure}

Morphologies were assigned to each SDSS object in our merging-pairs catalogue by a single classifier (DWD) working with consultation. We use four classifications for the merging galaxies: E, S, EU and SU. The E and S classifications respectively label those galaxies which are clearly elliptical and spiral by morphology. No appeal to colour should be necessary. The EU and SU are those ellipticals and spirals about which we are `unsure,' in other words, this is our best guess.\footnote{We refer to the set of merging galaxies with elliptical classifications as `E+EU' galaxies, etc.} The `unsure' morphologies are usually more distant objects whose image resolution is too poor to distinguish features like spiral arms. Choosing between EU and SU can be very difficult and is based mostly on apparent surface-brightness profile and, in very difficult cases, on colour. (See Figure \ref{morphChoice} for examples and further details of our decision-making criteria.)

A simple means to select morphology in SDSS is via the SDSS {\texttt fracdev} parameter measured in the $r$-band that ranges from 0 to 1. This is a measure of the goodness-of-fit of a galaxy's surface-brightness to a de Vaucouleur profile. Ellipticals tend to have a {\texttt fracdev} $\sim1$ whereas spirals have a wide distribution. We find that our morphological categories S, SU, EU and E have mean {\texttt fracdev} values of 0.48, 0.85, 0.93 and 0.94 respectively fitting qualitatively with expectation. The high value of 0.85 for the SU category suggests there is contamination by bulge dominated discs or ellipticals as discussed in \S \ref{s2e}. This is expected since distant, poorly-resolved spirals can look like ellipticals \citep{bamford}. To avoid such contamination we can restrict ourselves at anytime to using only the `sure' morphologies.

\section{The Merger Fraction of the Local Universe}

\label{merger_fraction}
Our large merging-pairs catalogue and $f_m$ values given by GZ for $893,292$ spectral objects enables us to address two distinct and important questions which are highly relevant to the field of galaxy evolution:

\begin{enumerate}

\item what is the {\bf fraction} of merging galaxies in the local universe? and

\item what is the {\bf ratio} of spirals to ellipticals ($N_s/N_e$) in mergers in the local universe?

\end{enumerate}

\subsection{Estimating the Merger Fraction (I)}
\label{aaa}

Our merger location technique is different to those before it and is suited therefore to a different procedure for finding a meaningful `merger fraction' which we make explicit here. First, we must find the fraction of volume-limited galaxies in the local universe currently in a `strongly-perturbed' state. To accomplish this we use the subset of our 893,292 GZ spectral objects that are members of the Main-Galaxy-Spectral sample (MGS; \citealt{strauss}) in SDSS and estimate what fraction of them are `strongly-perturbed.' By using only objects with spectra we are able to volume-limit the sample (by the usual $M_r<-20.55$ constraint) which is necessary for a meaningful merger fraction.

 \begin{center}

 \begin{equation} f_{mgs}=\frac{\sum \left[\begin{array}{c}\mbox{`Strongly-perturbed' volume-limited}\\\mbox{MGS spectral objects in SDSS}\end{array}\right]}{\sum \left[\begin{array}{c}\mbox{All volume-limited}\\\mbox{MGS spectral objects in SDSS}\end{array}\right]}.
\label{fraction}
\end{equation}
\end{center}

The use of only those spectral objects which are in the MGS leads to a good approximation for the {\it real} merger fraction, $f_{real}$, so long as an appropriate factor $C$ is applied to correct for spectroscopic incompleteness, i.e. $f_{real}=Cf_{mgs}$ where we estimate $C\sim1.5$. Justification for this claim with a brief discussion of the MGS and spectroscopic targeting in SDSS is given in Appendix A. 

The merging-pairs catalogue produced in \S \ref{sec2} is large but incomplete since it is constructed only from systems with $f_m>0.4$. Finding $f_{mgs}$ therefore requires extrapolation into the $0<f_m<0.4$ region. To find the numerator of (\ref{fraction}), we therefore took volume-limited samples of 100 GZ-spectral objects from the bins $0<f_m\leq0.05$, $0.05<f_m\leq0.20$ and $0.20<f_m\leq0.40$ which are also in the MGS. By visually inspecting each set of 100 galaxies {\it twice} (the first time round we made our decisions very conservatively, the second time round very liberally) we obtained estimates for the percentages of `strongly-perturbed' galaxies within these bins. These were: $0-2\%$, $18-34\%$ and $50-59\%$ for the $0.0<f_m\leq0.05$, $0.05<f_m\leq0.20$ and $0.20<f_m\leq0.40$ bins respectively.\footnote{We also double checked our results with an additional interface on the world-wide web designed to re-examine all objects within the range $0.2<f_m<0.4$. After a few months, enough users had re-classified these images of interest in order that a new-weighted-merger-vote fraction, $f_m^{\prime}$, could be calculated for each image based upon clicks of the `merger,' `not-merger' and `don't know' buttons. By then volume-limiting all spectral objects from this sample as before with $0.005<z<0.1$ and $M_r<-20.55$ we are able to filter out more actual `strongly-perturbed' systems. However, we again needed to decide a cutoff for the vote fraction, $f_m^{\prime}$. We examined twenty sets of ten images across the entire $f_m^{\prime}$ range and estimated how many of these were genuine `strongly-perturbed' systems. We found that $\ga80\%$ of images were actual `strongly-perturbed' systems for $f_m^{\prime}>0.6$ and, following \citet{lintott1}, made this number the cut off. The fraction of objects with $f_m^{\prime}>0.6$ is $\sim53\%$ which is in good agreement with our estimate that $50-59\%$ of objects in this range are `strongly perturbed.'} There are, of course, many more objects in these bins than in those with $f_m>0.4$ (see Figure \ref{votes_40_100}) and so end up contributing to at least two thirds of all `strongly-perturbed' systems. We assumed that no GZ object with $f_m=0$ would be `strongly-perturbed.' 

Applying these estimated fractions to the total number of MGS objects in each bin and adding the `strongly-perturbed' and volume-limited MGS objects from the 4198 GZ objects with $f_m>0.4$ gives us the numerator for equation (\ref{fraction}). The denominator of equation (\ref{fraction}) is found to be 157,801. Dividing these gives the fraction $f_{mgs}=4-6\%$. We sum up our result for {\color{black} $f_{real}$} in the following way:

\begin{quote} {\bf The Merger Fraction (Weak Statement)} \\
$\sim4-6\times C\% $ of all galaxies in a volume-limited sample ($M_r<-20.55$) in the local universe ($0.005<z<0.1$) are `strongly-perturbed.' 
\end{quote}

We call this the weak statement because of the subjective nature of deciding whether or not a system is `strongly-perturbed' or not. The error in this percentage arises from the upper and lower bounds estimated for the three samples of 100 images (before the correction factor $C \sim 1.5$ is applied). In \S \ref{aaa2} we use the 3003 systems from the merging-pairs catalogue (plus a few more plausible assumptions) to offer a stronger statement giving the fraction of {\it major} mergers in the local universe.

\subsection{The Stellar Masses of Merging Galaxies}
\label{mass}
\begin{figure}
	\includegraphics[width=84mm]{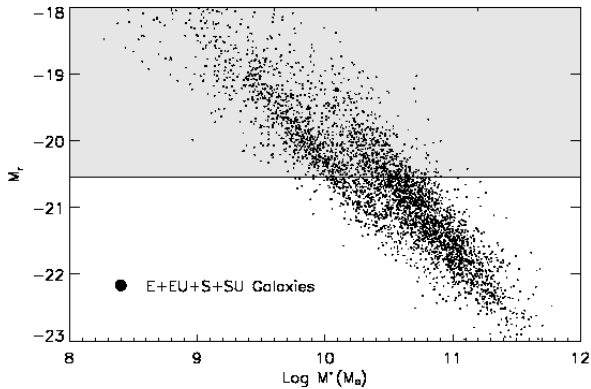}
	\caption[Figure 11]{Relation between stellar mass and absolute magnitude in the $r$-band. The solid horizontal line at $M_r=-20.55$ corresponds to the cut used to obtain our volume-limited samples. This cut removes virtually all galaxies with mass $\mbox{M}<10^{10}\mbox{M}_{\astrosun}$ as well as galaxies up to a mass of $M\sim7\times10^{10}\mbox{M}_{\astrosun}$.}	
	\label{scatter_kmass}
\end{figure}

Of the `strongly-perturbed' systems, we need to find what subset comprise major mergers. For this study, we define a major merger to be two merging galaxies of stellar masses M$_1^*$ and M$_2^*$ where $1/3<\mbox{M}^*_1/\mbox{M}^*_2<3$ and so, to estimate the subset of major mergers, we need to calculate stellar masses for the galaxies in the merging-pairs catalogue.

We do this by fitting the SDSS photometry for each object in our catalogue to a library of photometries produced by a variety of two-component star formation histories. The approach is similar to that of \citet{schawinski2}
except that we do not use the information contained in the stellar absorption indices. The library SEDs are generated using the Maraston (1998, 2005) stellar models. Both components have stellar populations with variable age with fixed solar metallicity and Salpeter IMF \citep{salpeter}. The first (older) burst is a simple stellar population (SSP), the second (more recent) burst is modeled by an exponential with variable e-folding time. The purpose of the varying e-folding times is to account for galaxies with extended star formation histories. This is especially important for mergers which are likely to have undergone recent star-formation episodes.  Dust is implemented using a Calzetti et al. law \citep{calzetti} that is free to vary for E(B-V) over 0 to 0.6. It should be noted that mergers, by their very nature, are prone to mix and overlap and this could lead to additional reddening.   

It is important that our cut for the merger fraction be based upon magnitude and not mass since it is the photometric brightness of the object that determines whether or not it will end up in the MGS (which is integral to our defintion of the merger fraction). In Figure \ref{scatter_kmass} we examine the effect of imposing our volume-limiting cut on the mass distributions. Mass scales with brightness though the scatter is substantial such that the magnitude limit $M_r<-20.55$ removes all galaxies with $<10^{10}\mbox{M}_{\astrosun}$ and some as massive as $\sim7\times10^{10}\mbox{M}_{\astrosun}$. It is therefore undesirable to estimate the merger fraction using a mass cut (instead of a magnitude cut) for this redshift range since one would need to cut at no less than $\sim7\times10^{10}\mbox{M}_{\astrosun}$ to ensure completeness and this would greatly reduce the sample size. It would also make a calculation of the merger fraction impractical since we would then need masses calculated for the entire MGS sample to get the denominator of equation (\ref{fraction}) and this would be an enormous computational task. In D09b we examine the distributions of these stellar masses in more detail and in comparison to a control sample.  

\begin{figure}
	\includegraphics[width=84mm]{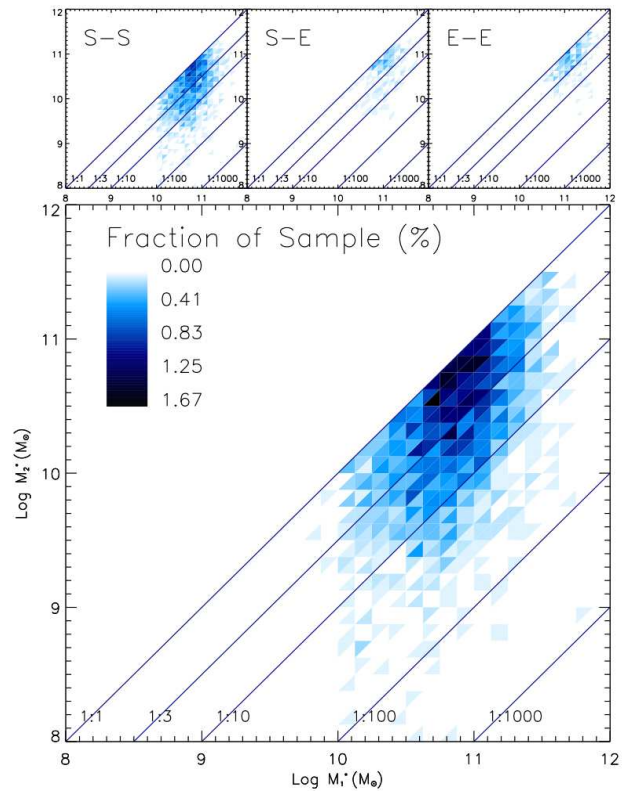}
	\caption[Figure 12]{This figure illustrates the occupancy of regions in mass-mass space for all the merging-pairs where the {\it more massive} galaxy has $M_r<-20.55$. Points between the $1:1$ and $1:3$ lines are major mergers. We only assume completeness within this strip in this study. The boxes in the upper panel divide the sample by morphology.}	
	\label{density}
\end{figure}

\subsection{Estimating the Merger Fraction (II)}
\label{aaa2}

We now apply the information obtained in \S \ref{aaa} (the fraction of `strongly-perturbed' galaxies) with that obtained in \S \ref{mass} (stellar-mass estimates for objects in the merging-pairs catalogue) in order to estimate the fraction of major mergers in a volume-limited sample ($M_r<-20.55$) in the local universe ($0.005<z<0.1$).

Figure \ref{density} illustrates the occupation of the mass space of the merging pairs in our catalogue with the magnitude limit imposed on the {\it more massive} galaxy in the pair.\footnote{{\color{black}As a reminder, the catalogue of 3003 merging pairs is {\it not} volume-limited, i.e. there is no formal constraint for either object in the pair to have a minimum absolute brightness. This means that in order to get a sample that is complete over the redshift range in use, we need to impose an absolute magnitude cut to all systems therein ($M_r<-20.55$ since, at $z=0.1$, this will select all galaxies with $r<17.77$, the minimum brightness needed to be designated as a spectral target). However, for mergers where we have two objects, one can choose to impose this absolute-magnitude constraint {\it either} on both objects in the pair {\it or} only on the largest/brightest object. We choose the latter option for this particular graph since it is more inclusive of minor-mergers.}} This criterion allows us to view both major and minor mergers. We find that $\sim50\%$ of these points lie within the major-merger strip. Estimation of the number of minor mergers within our volume-limited ranges is difficult since their completeness rapidly diminishes as $\mbox{M}^*_1/\mbox{M}^*_2$ increases.\footnote{Mergers between systems with $\mbox{M}^*_1/\mbox{M}^*_2>1000$ are of course abundant in the universe (every galaxy is merging with {\it something} small) but will not, by their very nature, get spotted in a merger study. We would therefore not be able to derive the abundance of minor-mergers in the local universe here with much confidence.} 

However, we can plausibly assume completeness for major mergers since, by the nature of the images they produce, they are easy to spot. Knowledge of the stellar-masses thus allows us to estimate what subset of the `strongly-perturbed' MGS spectral objects in the local universe which we found in \S \ref{aaa} are major mergers by the technical definition of $1/3 < \mbox{M}^*_1/\mbox{M}^*_2<3$.\footnote{As opposed to a purely subjective decision based upon visual inspection of the image.} 

To get a merger fraction, we again need to apply our analysis soley to volume-limited objects within the MGS.\footnote{Since we use the MGS to establish the denominator of equation (\ref{fraction}).} Of the 3864 spectral objects in our 3003 binary-merger pairs, 2306 have $M_r<-20.55$. Of these, 1243 are in major mergers leaving 1063 in minor mergers. We therefore find that $1243/2306\sim54\%$ of the volume-limited objects with spectra taken from our catalogue (which all have $f_m>0.4$ according to how they were selected) are in major mergers.

\begin{figure}
	\includegraphics[width=80mm]{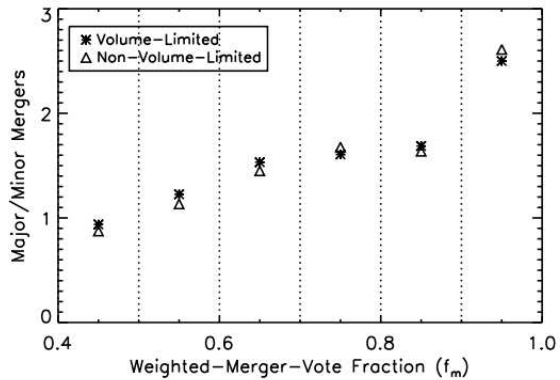}
	\caption[Figure 14]{The relationship of major-to-minor-mergers ratio for the merging-pairs catalogue over the $0.4<f_m<1.0$ range. The trend indicates that major mergers are more likely to get merger votes than minor mergers. Minor mergers only outnumber major mergers in the $0.4<f_m<0.5$ bin. The effect is similar for both volume- and non-volume limited samples.}	
	\label{maj2min}
\end{figure}

We can now estimate an upper-limit for the fraction of major mergers in the local universe by supposing that this fraction of major mergers ($\sim 54 \%$) from what was originally classified as a set of `strongly-perturbed' systems will be the same for {\it all} strongly-perturbed systems in the range $f_m<0.4$.\footnote{Recall that, when we estimated the fraction of `strongly-perturbed' systems as expressed in the Weak Statement, we needed to estimate how many MGS objects with $f_m<0.4$ would be classified as `strongly-perturbed.' Some of these will be major mergers, the rest minor. By assuming that the fraction of major mergers in our catalogue (with $f_m>0.4$) is the {\it same} for all the rest ($f_m<0.4$), we obtain an upper limit for how many major mergers are in the local universe.)} This is certain to be an {\it overestimate} since there are bound to be a higher portion of major mergers in systems with high $f_m$. The reason for this is that most `strongly-perturbed' systems obtain a low $f_m\sim0.2$ precisely because they are mostly `border-line' cases (i.e. perturbations caused by interactions or very minor mergers). Figure \ref{maj2min} confirms this expected relationship over the range of $f_m$ for our catalogue, that is, the ratio of major-to-minor mergers in our sample is seen to increase with $f_m$. In other words, users tended to more readily spot mergers involving galaxies of roughly equal mass. Applying this fraction of $54\%$ to the upper limit of our $f_{mgs}$ gives $0.54\times6\%=3.24\%$ which, being an overestimate, we can plausibly round down to $\sim3\%$.

We can also obtain a lower-limit to the major-merger fraction by supposing that the major mergers in our catalogue comprise {\it all} of the major mergers in the local universe. That is, we can suppose that there are {\it no} major mergers at all in the range $f_m<0.4$ for our volume-limited sample. This is of course an {\it underestimate} since there are bound to be at least some major mergers in, for example, the range $0.2<f_m<0.4$. Applying this assumption gives the lower limit of $1243/157,801\sim0.8\%$.\footnote{Where 157,801 is the denominator of (\ref{fraction}).} Again, since this is undoubtedly an underestimate, we can plausibly round this limit up to $\sim1\%$. We summarise this working in the following way.

\begin{quote} {\bf The Merger Fraction (Strong Statement)} \\
$\sim 1-3 \times C\%$ of all galaxies in a volume-limited sample ($M_r<-20.55$) in the local universe ($0.005<z<0.1$) are observed in a major merger.\footnote{Where we use the same corrective factor $C\sim 1.5$ (Appendix A).} 
\end{quote}

{\color{black} The large error obtained here arises due to the simplicity of the Galaxy Zoo interface. The recently released Galaxy Zoo Two project, which focuses on more specific questions (e.g. is the system `merging' or `interacting') and removes the dichotomy of describing {\it either} the morphology {\it or} the merger-likeness of the system, should lead to a much more exact figure for the merger fraction in the near future.} 

\subsection{Close-Pairs Comparison}
\label{close_pairs}

\begin{figure}
	\includegraphics[width=80mm]{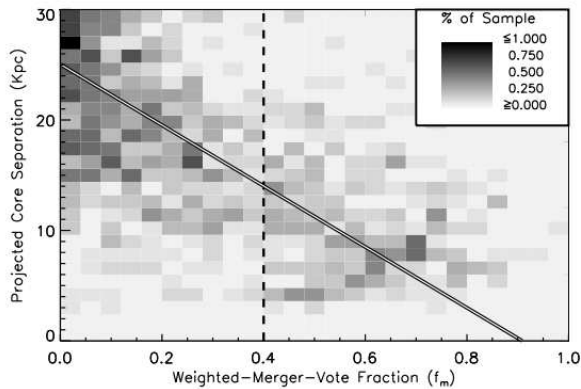}
	\caption[Figure 13b]{The distribution of volume-limited close-pairs objects in the MGS comparing $f_m$ and their projected separations. The vertical broken line marks the boundary for our merger catalogue at $f_m=0.4$. The solid diagonal line is a best-fit to the distribution of points in this $f_m$-separation space. The shading indicates the density of points in various regions of this space. GZ users were more likely to call a close-pairs object a merger as the projected core-separation decreased.}	
	\label{cpairs}
\end{figure}

Comparing the percentage in this strong statement with close-pairs studies is difficult because few of them present their results in terms of merger {\it fractions}. They also use different criteria for their volume-limiting bounds and for accounting for errors which are important in such studies (such as fiber collisions, false-pairs and non-gravitationally bound pairs). For example, \citet{patton2000} concluded that $\sim1.1\%$ of nearby galaxies with $-21<M_B<-18$ are undergoing a merger (but not necessarily a {\it major} merger by our definition). Our percentage is similar but the definitions are so diverse as to render comparison obsolete. 

A more recent study by \citet{patton2008} on SDSS close-pairs measures the number of `close-companions' per spectral object which satisfy three constraints: physical separation ($r_p$; $5<r_p<20h^{-1}$), rest-frame velocity ($\Delta v<500$kms$^{-1}$) and absolute magnitude difference ($\left| \Delta M_r \le0.753\right|$). Again though, there are significant differences between their samples and ours. Their magnitude limits are $-22<M_r<-18$ and they define a major merger by luminosity (at 1:2), not by mass as we do (at 1:3). More significantly, only $\sim$ half of the pairs satisfying their criteria are ``known to exhibit morphological signs of interactions'' whereas our sample is selected on the basis of visually-established interactions. So although the number of close-companions they calculate ($N_c\sim0.02$) is similar to our major-merger fraction ($\sim 1-3 \times C\%$) the differences between the two techniques make claims of corroboration difficult to substantiate.

We performed our own comparison of GZ with a close-pairs catalogue of all SDSS spectral objects within a projected separation of $30$kpc of each other and a line-of-sight velocity difference of $<500$kms$^{-1}$. For a consistent comparison with our catalogued mergers, we examine only those objects which are within the volume-limited boundaries ($0.005<z<0.1; M_r<-20.55$) and the MGS. This gives 2308 individual close-pair objects. Many of these will be part of a false-pair arising from the automated deblending of a single galaxy into two or more spectral targets (see \S \ref{cps}). Some pairs will also appear well separated (relative to their size) and show no signs of interaction. We therefore visually examined all 2308 objects in order to determine which ones are in a `strongly-perturbed' state brought about by the galaxy represented by the other spectral object in the close pair. This led to the removal of 654 ($\sim28\%$) of the close-pairs objects in our volume-limited MGS sample.

The 1654 remaining spectral objects are all GZ objects and so we can examine how users voted for them. We found that a significant portion ($\sim64\%$) of these close-pair spectral objects which {\it are} in `strongly-perturbed' systems (by our reckoning) had $f_m<0.4$ and were therefore excluded from our catalogue. However, we also found that there is a strong correlation between $f_m$ for close-pairs objects and their {\it projected separation} such that the further apart close-pair objects are, the less likely users were to label it a merger {\color{black}(see Figure \ref{cpairs})}. This means that the GZ technique selects systems which are generally more advanced in the merger process and therefore undergoing more extensive interactions than a typical system in close-pairs studies. Only $\sim600$ of these objects are in our catalogue of $\sim3000$ pairs which fits with the claim that only $\sim20\%$ of (relatively advanced) merging systems are spectral pairs in SDSS, the rest being systems with only a single spectral object within our magnitude-limited volume.
 
We conclude that, like the CAS and GM$_{20}$ techniques, the GZ method for detecting mergers is sensitive to {\it different} stages of a merger compared to the close-pairs technique (\citealt{lotz3}).
 
\subsection{Estimating the Spiral to Elliptical Ratio in Merging Systems}
\label{s2e}

\subsubsection{Uncertainties and User Bias}

Having established an estimate of what fraction of galaxies in the local universe are merging, we now turn to the more difficult task of estimating the spiral-to-elliptical ratio of galaxies that are merging ($N_s/N_e$).\footnote{Where $N_s$ and $N_e$ are the number of galaxies in a given sample of mergers that are spiral and elliptical respectively.} In addition to the uncertainties associated with identifying actual mergers, there are now the uncertainties associated with varying image resolution leading to `sure' and `unsure' morphologies (see \S \ref{assign_morph}). \citet{bamford} have studied extensively the dependence of GZ morphological classifications on redshift and found that the proportion of elliptical classifications increases with $z$. The same problem arose for our classifications: over the range $0.005<z<0.03$ the fraction of `unsure' morphologies (EU, SU) to `sure' morphologies (E, S) is $\sim 10\%$ but this fraction rises to $\sim25\%$ as we vary over $0.03\rightarrow0.1$. In short, the more distant a galaxy is the more `featureless' it appears bringing about a visual misclassification that inflates the recorded number of ellipticals. For details of the quantification and correction to this effect, see \citet{bamford}.

\begin{figure}
	\includegraphics[width=84mm]{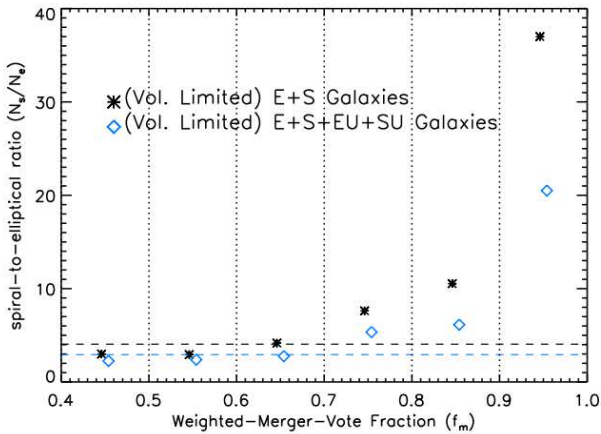}
	\caption[Figure 9]{An examination of the relation between morphology and $f_m$. We magnitude limit our sample ($M_r<-20.55$) and measure the ratios of spiral to elliptical morphologies ($N_s/N_e$) of mergers for bins shown along the $f_m$ axis. As $f_m\rightarrow1.0$ a strong bias is seen for users to flag mergers involving {\it spirals}. $N_s/N_e$ appears to converge to roughly a constant as $f_m\rightarrow0.4$. The broken horizontal lines are the mean $N_s/N_e$ values for the two samples over the whole range $0.4<f_m\le1.0$.}
	\label{voteEvolve}
\end{figure}

A further unknown peculiar to GZ is mass-user psychology. A natural concern of ours was that users would be more likely to call a system a `merger' if it involves two spirals (which generally look more dramatic) than if it involves two ellipticals. We therefore examine how the ratio of spirals to ellipticals in our merging-pairs catalogue varies with $f_m$ as shown in Figure \ref{voteEvolve}. Whether we use just the `sure' morphologies (E+S sets) or whether we include the `unsure' morphologies (E+S+EU+SU sets), the $N_s/N_e$ ratios appear to follow curves that start high for $f_m\sim1.0$ and decay towards a constant as $f_m\rightarrow0.4$. This confirmed the suspicion that the systems compelling users to click the merger button the most tend to involve spirals. However, the majority of mergers are located in the bins where the ratios level off at roughly a constant ($\sim 0.4<f_m<0.6$). Inclusion of the EU and SU categories slightly decreases $N_s/N_e$ as expected (some `unsures' are really spirals but get mistaken for ellipticals). 

The horizontal lines of Figure \ref{voteEvolve} indicate the mean $N_s/N_e$ ratios for the E+S and E+S+EU+SU samples over the whole range $0.4<f_m<1.0$. These $N_s/N_e$ means are $\sim5$ and $\sim3$ respectively. The true mean over this range must surely be $N_s/N_e >3$ therefore since the `unsure' morphological categories deflate $N_s/N_e$ through false inclusion of ellipticals as discussed earlier. 

\subsubsection{Towards an $N_s/N_e$ Estimate for Merging Galaxies}

We proceed now by extrapolating this value $N_s/N_e >3$ (found for the range $f_m>0.4$) to the entire range of $f_m$, that is, we assume $N_s/N_e >3$ for all galaxies in mergers for our volume-limited ranges. We compare this ratio with the global spiral-to-elliptical ratio for all galaxies in our redshift range determined using the corrections of \citet{bamford} to debias the high occurrence of ellipticals with increasing redshift. This debiasing leads to the estimate that there are $\sim 3:2$ spirals to ellipticals for all galaxies with $M_r<-20.55$ in our redshift range. Our extrapolated estimate of $N_s/N_e >3$ for $f_m>0.4$ is significantly higher in comparison, i.e. our extrapolation would suggest that spirals feature in mergers roughly twice as often as they should if selected randomly from the global population. 

To test the accuracy of this extrapolation we examined the morphologies of the same 59 systems from the 100 images of $0.2<f_m \leq 0.4$ (used in \S \ref{aaa}) that were deemed to be `strongly-perturbed.' We found that 42/59 of these `strongly-perturbed' galaxies were either S or SU and the remaining E or EU. Taking $N_s/N_e\sim1.5$ for the global estimate to give null binomial probabilities $p(spiral)=0.6$ and $p(elliptical)=0.4$, the expected outcome of 59 galaxy morphologies would be $35.4\pm\sigma$ spirals with standard deviation $\sigma\sim3.8$. The observed number, 42, is just within two standard deviations of the expected value. This observation therefore supports the claim that, even in the range $0.2<f_m<0.4$, $N_s/N_e$ in mergers is higher than the global mean. This also does not take into account the fact that our observations are still biased by inclusion of `unsure' morphologies (which inflates the number of ellipticals) whereas the estimate $N_s/N_e\sim1.5$ for the global population has been debiased.  


To summarise, we find that $N_s/N_e>3$ over the range of our merging-pairs catalogue ($f_m>0.4$). This is at least twice the global ratio ($N_s/N_e\sim1.5$). There is no evidence to suggest that mergers in the range $f_m<0.4$ will compensate this effect with an excess of ellipticals realtive to their global population. The high $N_s/N_e$ ratio is especially likely to stand up to scrutiny for {\it major}-mergers since they have been shown to favour higher values of $f_m$ (see Figure \ref{maj2min}) where the spiral excess has been robustly confirmed. 

\subsubsection{Implications of a High $N_s/N_e$ Ratio in Mergers}

We now discuss possible reasons for a high $N_s/N_e$ ratio in mergers. It is well established that spirals tend to occupy less-dense environments than ellipticals (verified as early as \citealt{oemler} and as recently as, for example, \citealt{ball}). The discrepancy might therefore be the result of a preference for mergers to occur in field environments where spirals are more populous relative to the global population. Alternatively though, the disproportionate number of spirals in our `snapshot' of the local universe could indicate that the time-scales over which spiral galaxies remain detectable in mergers exceeds that of ellipticals. 

Studies have been carried out to estimate the time-scales of mergers using dynamical-friction arguments (e.g. \citealp{con2006}) and simulations (e.g. \citealp{bell}; \citealp{con2006}; \citealp{lotz3}). \citet{lotz3}, in particular, focus on this question and find that the gas-fraction of galaxies in mergers is one of several factors determining the time-scale of detectability. One such simulation comparing two equal-mass mergers with different gas-fractions showed that the system with the higher gas-fraction ($f_{gas}\sim50\%$) remained `morphologically disturbed' (i.e. flagged as a merger by the standard criteria of G, $\mathrm{M}_{20}$, C, A combinations) for $2-4$ times longer than the system with a lower gas-fraction ($f_{gas}\sim20\%$). 

In practise, any variation in $N_s/N_e$ for merging galaxies will depend on a combination of these two explanations since, {\it a priori}, it seems certain that the merger frequency will have {\it some} dependence on environment and that the merger-time-scale of detectability will have {\it some} dependence on the internal characteristics that distinguish spirals from ellipticals (such as gas-content and overall mass). We examine these two effects in more detail in the companion paper D09b and conclude that, in fact, the latter explanation seems to be the most likely, that is, that mergers involving spirals remain detectable for longer times than mergers with involving ellipticals.


\section{Summary and Discussion}
\label{conc}

Galaxy Zoo is a new and powerful strategy for locating mergers. The technique is similar in its effect to the CAS and GM$_{20}$ methods in that it converts images to numbers that provide a measure of how `merger-like' a galaxy is (in this case the weighted-merger-vote fraction $f_m\in[0,1]$). The method is highly apt at locating mergers because $f_m$ is the averaged product of human minds (which are highly adept at pattern recognition) and is therefore extremely sensitive to details while doing away with the major programming challenges associated with automated methods. It is also easier to partition merger-spaces in this method since $f_m$ is a single parameter unlike CAS and GM$_{20}$ which demand more fine-tuning. The technique is also not limited to objects with spectra as the close-pairs method is (by definition) and we find in fact that only $\sim20\%$ of our merging systems have spectra on both galaxies. This is mostly due to the large fraction of the SDSS sky that suffers from fiber-collisions ($\sim 70\%$; \citealt{strauss}, \citealt{blanton2}). It also means that the method is effective at the broader task of finding `strongly-perturbed' systems including minor mergers. The drawbacks to GZ are time and repeatability. The results presented in this paper are derived from about six months of web activity. 

It is worth emphasising that the results of this paper are entirely derived from the pressing of a single button labeled `merger.' The simplicity of this interface lead to an unfortunate dichotomy whereby users were often unsure whether to emphasise the merger aspect or the morphological aspect of a given system. Now that the competence and eagerness of the public to assist in extragalactic astronomy is known and tested, we expect greatly improved results with the Galaxy Zoo Two project - a new development that will enable finer classification of SDSS objects as well as higher redshift surveys.

Galaxy Zoo has already yielded rich results with the initial merger catalogue presented here containing 3003 merging-pairs in mergers in the range $0.005<z<0.1$ created from GZ objects with $f_m>0.4$. Each has been been visually-inspected by fifty or so Galaxy-Zoo users and by one of the authors (DWD). We believe that it is the largest of its kind. Completeness, however, remains an issue as users and experts are only as accurate as the image quality allows. As redshift increases, the reduced image quality makes it difficult to identify galactic projections. Also, there will always be the problem for any merger-detection method in deciding whether or not a galaxy perturbation (such as an extended tidal tail) is great enough to warrant the term `merger.' For these two reasons, there is a large number of `border-line' cases in the range $0<f_m<0.4$ which are responsible for the error in our estimate of the merger fraction in the local universe.

To obtain this merger fraction, we first estimated the number of `strongly-perturbed' galaxies with spectra in bins over the range $0<f_m<0.4$ plus those in our catalogue. Dividing this number by the total number of volume-limited spectral objects lead us to estimate {\color{black} that $f_{mgs} \sim 4-6\%$. With the correction factor $C\sim 1.5$ converting $f_{mgs}$ to $f_{real}$ estimated in Appendix A, one can say that $\sim4-6 \times C \%$ of all volume-limited galaxies ($0.005<z<0.1$ and $M_r<-20.55$) are `strongly-perturbed.' We expanded upon this statement by estimating what subset of `strongly-perturbed' volume-limited spectral galaxy objects are major mergers in the local universe. This led to the stronger statement that: }

\begin{quote}$\sim1-3 \times C \%$ of volume-limited galaxies in the local universe ($M_r<-20.55, 0.005<f_m<0.1$) are observed in major mergers.
\end{quote}

We also estimated the ratio of spirals to ellipticals for merging galaxies with $M_r<-20.55$. The $N_s/N_e$ ratio for our catalogue was highly in favour of spirals by $\sim 3:1$. We examined 100 systems in the range $0.2<f_m<0.4$ and found a similar result. This is large compared to the global $N_s/N_e$ ratio ($\sim 3:2$) for this magnitude-limited volume. 

We therefore concluded that more spirals are seen to be merging than ellipticals, perhaps by as much as factor $\sim2$. We then discussed possible reasons for this observed spiral excess suggesting that either (1) mergers tend to occupy environments that favour spirals, or (2) the time-scale for a merger to reach a relaxed state (i.e. the time over which it is detectable) is longer for spirals than for ellipticals.\footnote{The probability, $p_{o}$, of observing a galaxy merging at any general time is proportional to the probability of its merging at that time, $p_m$, multiplied by the time-scale, $\tau$, over which it is detectable, $p_{o}\propto p_m\tau$. The probability of merging should only depend on the system's mass and environment ($p_m=p_m(\mbox{M},\rho)$) whereas the time-scale of detectability, $\tau$, must be assumed {\it a priori} to depend on the internal properties of the merging galaxy such as gas content and mass distribution. Our high $N_s/N_e$ ratio therefore suggests that either $p_m(\mbox{M},\rho)$ or $\tau$ is large for spirals compared to ellipticals.} 

We disentangle these effects in the companion paper D09b by exploring the role of the environment and the internal properties of this merger sample. We find in fact that mergers occupy the same (if not slightly {\it denser}) environments as a randomly selected control sample of galaxies. We conclude there that the best explanation of an apparently high spiral-to-elliptical ratio in mergers, as we find here, must be due to varying time scales of detectability. The properties of spirals and ellipticals that affect merger detactability are therefore an important issue that the Galaxy Zoo catalogue can help us understand. 

\section{Acknowledgments}
\label{ack}

D. W. Darg acknowledges funding from the John Templeton Foundation. S. Kaviraj acknowledges a Research Fellowship from the Royal Commission for the Exhibition of 1851 (from Oct 2008), a Leverhulme Early-Career Fellowship (till Oct 2008), a Senior Research Fellowship from Worcester College, Oxford and support from the BIPAC institute, Oxford. C. J. Lintott acknowledges funding from The Levehulme Trust and the STFC Science in Society Program. K. Schawinski was supported by the Henry Skynner Junior Research Fellowship at Balliol College, Oxford.

Funding for the SDSS and SDSS-II has been provided
by the Alfred P. Sloan Foundation, the Participating Institutions,
the National Science Foundation, the U.S. Department
of Energy, the National Aeronautics and Space
Administration, the Japanese Monbukagakusho, the Max
Planck Society, and the Higher Education Funding Council
for England. The SDSS Web Site is http://www.sdss.org/.
The SDSS is managed by the Astrophysical Research Consortium for the Participating Institutions. The Participating
Institutions are the American Museum of Natural
History, Astrophysical Institute Potsdam, University of
Basel, University of Cambridge, Case Western Reserve University,
University of Chicago, Drexel University, Fermilab,
the Institute for Advanced Study, the Japan Participation
Group, Johns Hopkins University, the Joint Institute for
Nuclear Astrophysics, the Kavli Institute for Particle Astrophysics
and Cosmology, the Korean Scientist Group, the
Chinese Academy of Sciences (LAMOST), Los Alamos National
Laboratory, the Max-Planck-Institute for Astronomy
(MPIA), the Max-Planck-Institute for Astrophysics (MPA),
New Mexico State University, Ohio State University, University
of Pittsburgh, University of Portsmouth, Princeton
University, the United States Naval Observatory, and the
University of Washington.


\appendix
\section{Estimating the Merger Fraction Using Spectral Objects}
\label{tiling}

\begin{figure}
	\includegraphics[width=70mm]{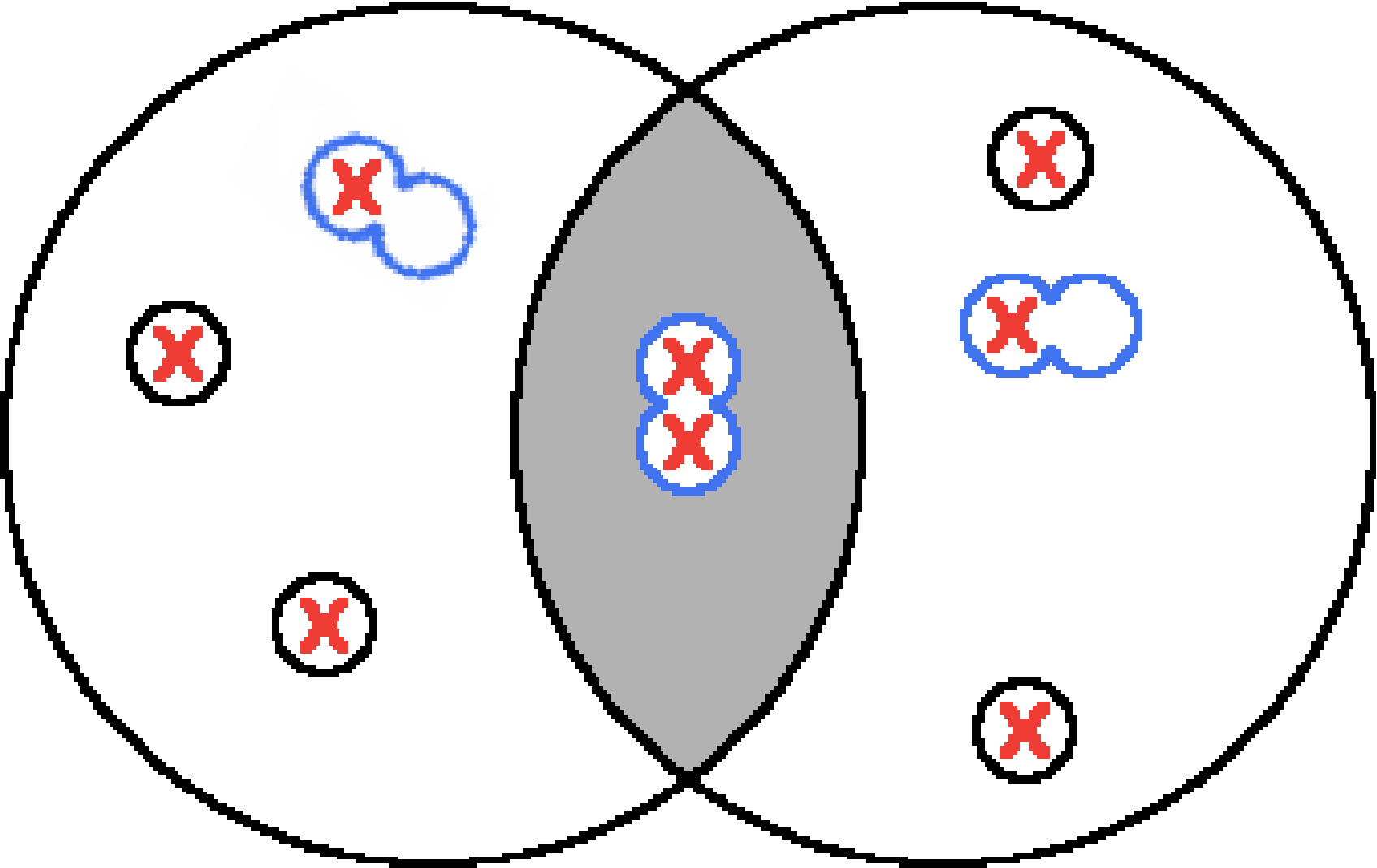}
	\caption[Figure App1]{An illustration of the SDSS spectroscopic tiling scheme. The overlap region shaded grey covers a fraction $k_{sky}$ of the area. Strongly-perturbed systems are coloured with blue outlines and single galaxies black. Those targets which receive spectra are marked with a red cross. Here we only consider systems where some strongly-perturbed spectral objects are merging with spectral targets.}	
	\label{tile}
\end{figure}

Here we clarify some of the subtle issues that arise when attempting to estimate the merger fraction of the universe. It was claimed in \S \ref{aaa} that one could use spectral objects alone to obtain an accurate estimate of the merger fraction. This is true even though a sizable number of galaxies in the universe are strongly-perturbed but do not have spectra due to fiber collisions (two spectral targets within $55''$ cannot both obtain spectra in regions with only a single tiling). In overlap regions, there is near spectral completeness for spectral targets (\citealt{strauss}; \citealt{blanton2}; see Figure \ref{tile}). 

In calculating the merger fraction, we only use spectral objects belonging to the Main-Galaxy-Spectroscopic sample (MGS) in SDSS. This is the collection of targets identified photometrically as galaxies (but not quasars or luminous red-galaxies; \citealp{stoughton}; \citealp{strauss}) with $r<17.77$ and is intended for statistical sampling of galaxies from a uniform population \citep{strauss}.

Figure \ref{tile} depicts a simple tiling scheme analogous to that employed by SDSS \citep{blanton2}. The shaded area represents the overlap which we say covers a fraction $k_{sky}$ of the tiled area (for SDSS $k_{sky}\sim30\%$). Two objects within $55''$ in non-overlap regions suffer from fiber-collision so that only one can get measured. Within overlap regions virtually all targets get measured due to multiple tiling opportunities. Altogether about $6\%$ of all MGS targets do not obtain spectral measurements due to fiber collisions \citep{strauss}. These are, of course, the objects most likely to be of interest in a mergers study though many might be projections, so one cannot assume they are part of a perturbed system.

We argue now that our estimate for the fraction of `strongly-perturbed' volume-limited MGS objects ($f_{mgs}$) is, upto a correction factor close to unity, a good approximation to the `real' fraction of volume-limited galaxies ($f_{real}$). For simplicity, let us begin with a sky as depicted in Figure \ref{tile} involving only galaxies which are bright enough to be spectral targets and where some of these might be in the process of a binary merger such that they are within $55''$ of another spectral target. We assume that mergers are randomly distributed with respect to tile-overlap regions. Let the number of single (i.e. non-perturbed) galaxies be $N_s$ and the number of binary-merger {\it pairs} be $N_m$. The merger fraction that we calculated, $f_{mgs}$, would therefore be

\begin{eqnarray}
f_{mgs} &=&\frac{2k_{sky}N_m + (1-k_{sky})N_m}{2k_{sky}N_m + (1-k_{sky})N_m+N_s}\\
	&=&\frac{N_m(1+k_{sky})}{N_m(1+k_{sky})+N_s}.
\label{a1}
\end{eqnarray}   

The real merger fraction, ($f_{real}$) should take account of everything regardless of overlap regions so that the fraction of strongly perturbed spectral targets is 

\begin{equation}
f_{real} = \frac{2N_m}{2N_m+N_s}.
\label{a2}
\end{equation}   
 
Now we let the ratio of binary-merger pairs $N_m$ to single galaxies in the local universe equal $p$. Taking the ratio of (\ref{a1}) and (\ref{a2}) and substituting $N_m=pN_s$ will give the corrective factor, $C$, relating our fraction to the real fraction:

\begin{equation}
C(k_{sky},p)=\frac{f_{real}}{f_{mgs}} = \frac{2(1+pk_{sky}+p)}{(2p+1)(1+k_{sky})}.
\label{a3}
\end{equation}   

When $k_{sky}=1$ (i.e. when the whole sky is an overlap region) we get $f_{real}=f_{mgs}$. For realistic values like $k_{sky}\sim30\%$, $p\sim5\%$ we get a ratio $C(k_{sky}=0.3,p=0.05)\sim1.5$ and we find that (\ref{a3}) is not very sensitive to changes in $p$. Multi-mergers have the affect of increasing $C$ but are rare enough to be ignored (only comprising $\sim 1\%$ of the merging-pairs catalogue) and so we can take our corrective factor to be $C=1.5$. 

\end{document}